\documentclass{article}

\usepackage{arxiv}
\usepackage{graphicx}
\graphicspath{ {./Figure/} }

\usepackage{xr}
\usepackage{url} 
\usepackage{float}
\usepackage{natbib}
\usepackage{lineno}
\usepackage{xspace}
\usepackage{upgreek}
\usepackage{siunitx}
\usepackage{makecell}
\usepackage{soul,color}
\usepackage{amssymb,amsmath}
\usepackage{multirow,booktabs}
\usepackage{hyperref}
\usepackage[nameinlink]{cleveref}

\graphicspath{ {./figures/} }

\makeatletter
\newcommand*{\addFileDependency}[1]{
  \typeout{(#1)}
  \@addtofilelist{#1}
  \IfFileExists{#1}{}{\typeout{No file #1.}}
}
\makeatother

\newcommand{\modelname}{Zeeman\xspace}

\title{\modelname: A Deep Learning Regional Atmospheric Chemistry Transport Model}

\author{
  Mijie Pang$^1$, Jianbing Jin$^{2,*}$, Arjo Segers$^3$, Hai Xiang Lin$^{1,4}$, Guoqiang Wang$^{5}$, Hong Liao$^{2}$, Wei Han$^{6*}$\\
  \And
  $^*$Corresponding authors: Jianbing Jin (\texttt{jianbing.jin@nuist.edu.cn}) and Wei Han (\texttt{hanwei@cma.gov.cn})\\
  \And
  $\quad$\\
  $^{1}$Delft Institute of Applied Mathematics, Delft University of Technology Delft, the Netherlands\\
  $^{2}$State Key Laboratory of Climate System Prediction and Risk Management, \\Jiangsu Key Laboratory of Atmospheric Environment Monitoring and Pollution Control, \\Jiangsu Collaborative Innovation Center of Atmospheric Environment and Equipment Technology,\\ School of Environmental Science and Engineering, \\Nanjing University of Information Science and Technology, Nanjing, Jiangsu, China\\
  $^{3}$Department of Climate, Air and Sustainability, TNO, Utrecht, the Netherlands\\
  $^{4}$Institute of Environment Sciences, Leiden University, Leiden, the Netherlands\\
  $^{5}$School of Mathematics, Physics and Statistics, Shanghai University of Engineering Science, Shanghai, China\\
  $^{6}$CMA Earth System Modeling and Prediction Centre, Chinese Meteorological Administration, Beijing, China
}

\begin{document}
\maketitle

\begin{abstract}
Atmospheric chemistry encapsulates the emission of various pollutants, the complex chemistry reactions, and the meteorology dominant transport, which form a dynamic system that governs air quality. While deep learning (DL) models have shown promise in capturing intricate patterns for forecasting individual atmospheric components—such as PM$_{2.5}$ and ozone — the critical interactions among multiple pollutants and the combined influence of emissions and meteorology are often overlook. This study introduces an advanced DL-based atmospheric chemistry transport model \modelname for multi-component atmospheric chemistry simulation. Leveraging an attention mechanism, our model effectively captures the nuanced relationships among these constituents. Performance metrics demonstrate that our approach rivals numerical models, offering an efficient solution for atmospheric chemistry. 
In the future, this model could be further integrated with data assimilation techniques to facilitate efficient and accurate atmospheric emission estimation and concentration forecast.
\end{abstract}

\section{Introduction}
Atmospheric chemistry, a field that intertwines the subtle dance of emissions, concentrations, and meteorological patterns, stands as a vivid illustration of the intricate web of interactions within our environment. At its core, this discipline explores how pollutants and natural substances are emitted into the atmosphere, transform through chemical reactions, and are transported by the wind and other atmospheric processes.
\citep{sokhi2022advances}. The complexity arises from the fact that these components do not operate in isolation; rather, they form a dynamic system where each element influences the others in profound ways. Emissions, originating from anthropogenic and natural sources, introduce diverse chemicals into the air, setting the stage for a series of chemical reactions \citep{xiao2021separating}. These reactions, in turn, are heavily influenced by the prevailing meteorological conditions \citep{qu2021study}, such as temperature and humidity \citep{su2020new}, which can either accelerate or decelerate reaction rates \citep{mu2018temperature}. Meanwhile, the concentration of these substances in the atmosphere, shaped by both emission rates and removal processes, plays a crucial role in determining the impact on air quality and climate \citep{raes2010atmospherica}. 
Understanding the interplay between emissions, concentrations, and meteorology is crucial for performing air quality forecasting and addressing pressing environmental challenges \citep{zhang2008onlinecoupleda,lu2019meteorology}.
Moreover, this understanding serves as the foundation for data assimilation and emission inversion, which are key focuses in atmospheric research \citep{zhu2023recent}.
While this interplay brings substantial computational demands and casts great challenges to efficiently consider the intricate dynamics of atmospheric chemistry \citep{feng2021wrfgc}. High demand of computation can be noticed especially when atmospheric chemistry is involved. \citep{grell2011integrated,gao2024intercomparison}.

Deep learning (DL) models have demonstrated remarkable capabilities in processing and analyzing large amounts of data \citep{yu2021deep}. These DL models can capture complex patterns and relationships within atmospheric conditions, providing precise and timely forecasts \citep{deburgh-day2023machine}. Numerous of researches have applied DL models to perform forecasts within different aspects of atmospheric chemistry \citep{zhang2022deep,mendez2023machine,tang2024review}.
\cite{zhang2020constructing} proposed a model based on an auto-encoder and bidirectional long-short term memory (Bi-LSTM) to predict the PM$_{2.5}$ concentration. PM$_{2.5}$ concentration, meteorological factors are used as inputs in this model.
\cite{cheng2022spatiotemporal} combined a generative adversarial network (GAN) with a variational autoencoder (VAE) to learn the relationship between meteorological factors and ozone. Long lead-time ozone predictions can be made by this model. 
\cite{lu2023highresolution} developed a space-time Light Gradient Boosting Machine (STLGB) model to estimate the spatial distribution of Non-methane volatile organic compounds (NMVOCs). The model incorporates NMVOCs station observations, satellite-derived emissions data, and meteorological information as inputs. The strong influence of emissions on NMVOCs estimation were emphasized. 
While these models exhibit relatively good performance in predictions, they are typically constrained to single component or combined indices, such as PM$_{2.5}$ and the Air Quality Index (AQI). Interplay between diverse components is often overlooked, limiting the models' comprehensiveness. For example, nitrogen oxides (NOx) and volatile organic compounds (VOCs) undergo a series of complex chemical reactions under sunlight, leading to the formation of O$_3$ \citep{george2015heterogeneous}. Precursor substances like sulfur oxides (SOx), nitrogen oxides (NOx), and ammonia (NH$_3$) can transform into secondary aerosols such as sulfates, nitrates, and ammonium salts which constitute PM$_{2.5}$ through various chemical reactions in the atmosphere \citep{peng2021explosive}. These components are closely related to each other, and their interactions are essential for building the atmospheric chemistry transport model. 
Furthermore, emissions and meteorology play an crucial role in determining the increase or decrease of concentrations. Relative humidity (RH) has a substantial impact on the evolution of secondary aerosol in the atmosphere \citep{ma2021effects}. Surface ozone can be effected by boundary layer height (blh) in a complex way \citep{zhang2023relationship}. On the other hand, emissions are the major source of air pollutions \citep{li2023emission}. While emissions are frequently overlooked when designing a deep learning based prediction model due to difficulties in acquiring high spatio-temporal resolution information. 
In general, meteorological fields and emissions are inherently integrated with atmospheric concentration levels \citep{wang2024quantitative}. To achieve an accurate air quality prediction, incorporation of both elements is necessary. Nevertheless, their complex dynamics also present significant challenges to modelling efforts.

In recent years, the integration of deep learning methodologies into weather prediction has inaugurated a new epoch characterized by unprecedented levels of accuracy and computational efficiency \citep{bi2023accurate,lam2023learning,chen2023fuxi}. These advancements, collectively termed artificial intelligence (AI)-based approaches, have enabled models to discern intricate patterns within atmospheric dynamics, thereby offering superior forecast precision and timeliness compared to conventional numerical weather prediction (NWP) techniques \citep{charlton-perez2024ai}.
This study introduces an advanced deep learning driven framework for atmospheric chemistry transport - \modelname. Our contributions are twofold:
Comprehensive Multi-Pollutant Simulation: We expand the scope of atmospheric chemistry components to encompass O$_{3}$, NH$_{3}$), NO$_{2}$, fine particulate matter (PM$_{2.5}$), and coarse particulate matter (PM$_{10}$). An detailed inventory of these components is provided in \cref{tab:dataset}. By employing an attention mechanism, our model adeptly captures the nuanced interactions among these constituents, achieving performance metrics that rival those of numerical model across main components.
Enhanced Framework via Integrated Data Streams: The efficacy of our model is further improved by the incorporation of hourly meteorological data alongside emissions. Given the pivotal role of both meteorological parameters and emission sources in determining atmospheric concentrations, this enriched dataset facilitates the generation of stable predictions with extended lead times. In addition, \modelname is a regional model whose forecast can be impacted by external inputs. Therefore, the boundary conditions are also considered in \modelname. 

While the study primarily focuses on designing a fast, surrogate model that approximates LOTOS-EUROS, its applications extend well beyond this. For example, \modelname can be off-line integrated into meteorological models to provide coupled forecasts. Data assimilation can be implemented using \modelname through variational or ensemble methods, providing a framework that delivers greater accuracy and efficiency \citep{zhong2024fuxiens,wang2024fourdimensional,li2024fuxien4dvar}.


This paper is organized as follows: \Cref{sect:dataset} introduces the dataset used to train the model.
\Cref{sect:method} introduces the overall architecture of \modelname and training details.
\Cref{sect:results} evaluates the performance of \modelname forecasts in terms of spatial, vertical and long-term forecasts. 
In the end, \Cref{sect:conclusion} concludes the paper and point out limitations and future work.

\section{Dataset}
\label{sect:dataset}
This section introduces all the data used to train the \modelname including 3D concentrations from LOTOS-EUROS simulations, emission, meteorological fields and boundary conditions. These emission, meteorology and boundary fields are also driving LOTOS-EUROS model to produce concentration simulations. The dataset spans the period from 2018 to 2022. The first four years (2018–2021) are used for training the model and the last year, 2022, is reserved for evaluating model performance.

\subsection{LOTOS-EUROS simulations}
We used the open-source CTM LOTOS-EUROS v2.2, a three-dimensional (3D) regional chemistry transport model for simulation of trace gases and aerosol concentrations \citep{manders2017curriculum,lotos-euros}. It has been wildly used for air quality researches and forecasting \citep{timmermans2017source,skoulidou2021evaluation,timmermans2022evaluation}. LOTOS-EUROS constitutes one of the state-of-the-art atmospheric chemistry models used by the Copernicus Atmosphere Monitoring Service \citep{colette2024copernicus} to provide daily forecasts of the main air pollutants, i.e. O$_{3}$, NO$_{2}$ and PM$_{10}$. 

\subsection{Simulated concentrations}
In this study, 3D concentration are generated by the LOTOS-EUROS for the main air pollutants.
These include trace gases such O$_3$, NOx, NH$_3$, carbon monoxide (CO), and NMVOC, and aerosols (particulate mater) of various types for size bins within a range of 0-10 $\si{\micro\meter}$. 
A full model configuration has 54 transported tracers.
For the \modelname model, a limited set of 17 tracers is used which are either one of the original tracers, or an accumulated tracer such as total PM$_{2.5}$ and PM$_{10}$ (all particulate matter with diameters less than 2.5 or 10 $\si{\micro\meter}$). 
The full list of selected concentrations is detailed in \cref{tab:dataset}.

\subsection{Model domain and resolution}
The model domain is limited to an area over Netherlands, illustrated in \cref{fig:city_map}.
Simulations in this domain are at a resolution of $0.1^\circ\times0.1^\circ$, 
which is the same as the operational resolution in the CAMS forecast.
For the chosen domain, the model simulates concentrations at 50 $\times$ 40 grid cells of approximately 10 x 5 km wide.
In the vertical, 11 layers are used that are a coarsening of the ECMWF operational meteorological data described below; the lowest layer is about 20 m thick, and the top is around 9 km.

\subsection{Boundary conditions}
For a regional atmospheric chemistry model, boundary conditions plays an important role especially after lone time forecasting \citep{qu2024effect}. Then boundary conditions of concentration is then included in Zeeman. The boundary conditions for the European run are obtained from the CAMS near-real-time global chemistry simulations \citep{peuch2022copernicus}.
The simulations from the European domain are used as boundary conditions for the high-resolution simulation at the target domain, where all 54 tracers of the full model are transferred.
The \modelname model is however trained using the boundary values for the 17 selected (accumulated) tracers, obtained from the 1-layer shell of grid cells around the target domain. Experiment on the necessity of including boundary conditions is shown in SI \cref{sect:si-boundary}.

\subsection{Emission}
This study employs the emission inventory from the Copernicus Atmosphere Monitoring Service REGional inventory (CAMS-REG) \citep{kuenen2021copernicus,kuenen2022camsregv4a}, a specialized dataset designed to facilitate air quality modelling. CAMS-REG provides anthropogenic emissions data for Europe, starting from 2020, at a spatial resolution of $\qty{0.05}{\degree}\times\qty{0.1}{\degree}$. The inventory encompasses a range of substances, including both air quality pollutants and greenhouse gases. For this work, we have selected the air pollutants, which includes emissions of nitrogen oxides (NOx), sulfur dioxide (SO$_2$), non-methane volatile organic compounds (NMVOCs), ammonia (NH$_3$), carbon monoxide (CO), components of particulate matter(PM$_{10}$), as well as methane (CH$_{4}$). A comprehensive list of the species utilized in this work is detailed in \cref{tab:dataset}. List of full names can be found in \cref{tab:names}. Note that the emissions at time $t$ is the averaged emissions between $t$ and $t+1$. This one-step ahead strategy is adopted help model better characterize the impact of emissions to the future concentration states.

\subsection{Meteorology data}
Meteoroloty data is from ECMWF operational forecasts over 0-12 hour. Three-dimensitional (3D) variables including temperature (t), pressure (p), wind speed (uv\_u, uv\_v) and relative humidity (rh) are used as the input of training. Besides, 2D information of boundary layer height (blh), rain is also included in the training. These data are spatial-temporal interpolated to model grids. Time $t$ of this data implies the instant values for the middle of the hour. 

\begin{table}[ht]
    \centering
    \caption{Overview of the dataset.}
    \label{tab:dataset}
    \resizebox{\linewidth}{!}{
        \begin{tabular}[c]{cccc}
        \toprule
            Category  & Dimension & Dimension type & Species \\
        \midrule
                Concentration & 17 × 11 × 40 × 50 & 3D & \makecell[c]{no2, o3, co, so2, nh3, nh4a, pan, so4a, no3a\_f, no3a\_c,\\ ec, pom, ppm, tnmvoc, tpm25, tpm10, tss} \\
            Meteorology3d & 5 × 11 × 40 × 50 & 3D & t, rh, u, v, p \\
                Meteorology2d & 2 × 1 × 40 × 50 & 2D & blh, rain \\
                Emission & 27 × 1 × 40 × 50 & 2D & \makecell[c]{no2, no, co , form, ald, par, ole, eth, tol, xyl, so4a\_f, so2,\\ ch4, nh3, iso, terp, ec\_f, ec\_c, pom\_f, pom\_c, ppm\_f,\\ ppm\_c, na\_ff, na\_f, na\_ccc, na\_cc, na\_c} \\
                Boundary & 17 x 11 x 2 x (40 + 50) & 3D & Same as concentration \\
        \bottomrule
        \end{tabular}
    }
\end{table}

\section{Methodology}
\label{sect:method}
This section introduces the overall architecture of the \modelname including the architecture and training details. And the forecast strategy of \modelname.

\subsection{Model architecture}
\begin{figure}
    \centering
    \includegraphics[width=0.9\linewidth]{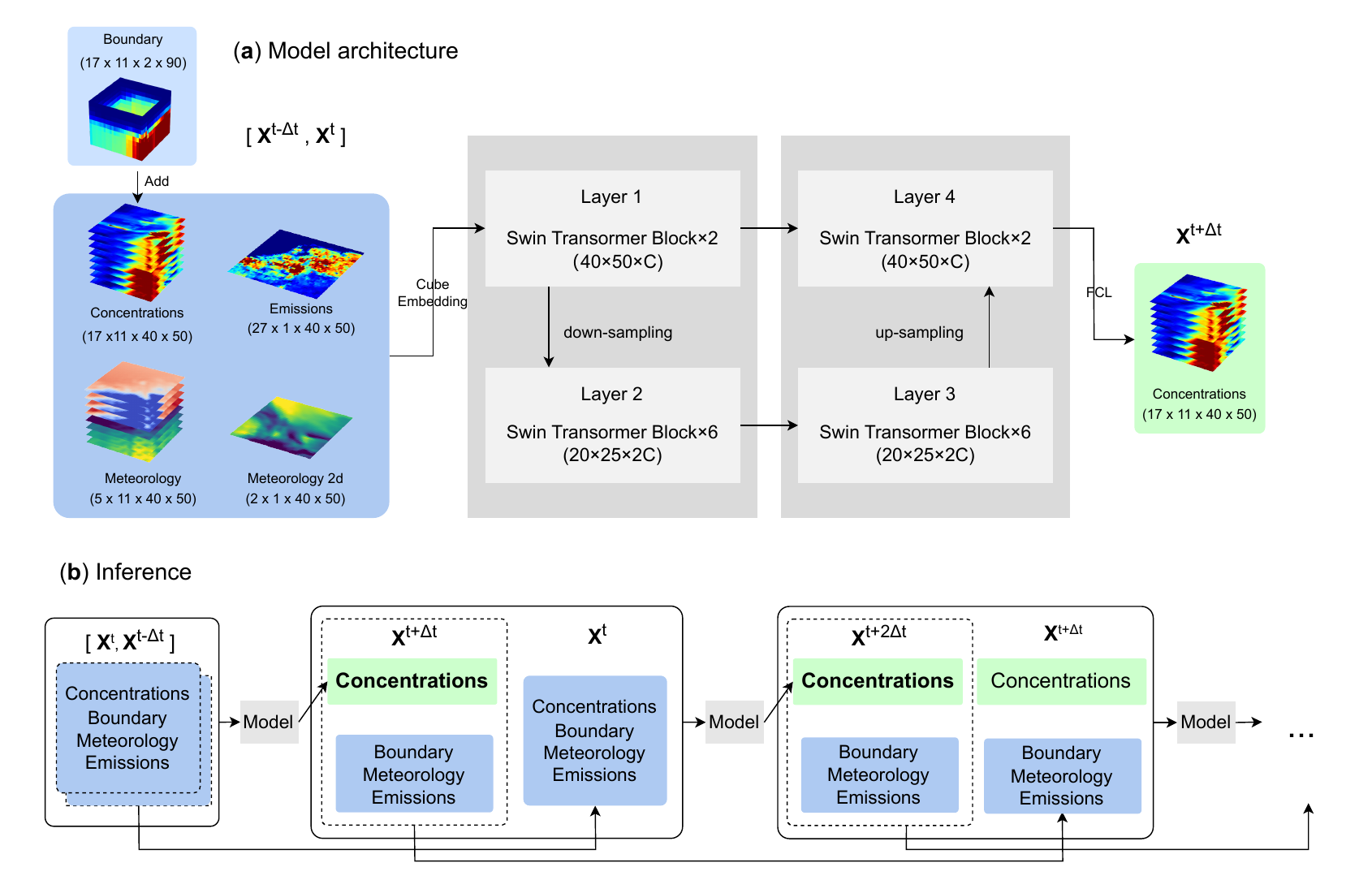}
    \caption{Overall architecture of \modelname. (\textbf{a}) the main components of \modelname: boundary enhancement, cube embedding, Swin-transformer, full connected layer. (\textbf{b}) the process of cyclic prediction based on \modelname.}
    \label{fig:architecture}
\end{figure}

The architecture of \modelname consists of four main components, which are illustrated in \cref{fig:architecture}: boundary enhancement, cube embedding, Transformer blocks, and a output layer. 

In the context of regional prediction models, boundary conditions play a crucial role, particularly for forecasts with longer lead times. Given that these models are predominantly influenced by wind patterns, there is a potential for concentrations outside the domain to be introduced into the area, which can result in underestimations of concentration if neglected. To address this issue, three-dimensional boundary values have been incorporated into the original concentration fields. This enhancement aims to provide a more accurate representation of concentration levels by accounting for external influences that may affect the region over longer forecast periods.
Afterwards, the space-time cube embedding is applied. The input data combines 3D, multi-component variables and creates a data cube with dimensions of C × H × W, where C, H, and W represent the total number of input channels, latitude and longitude grid points, respectively. H is 40 and W is 50 here. C, the channel width of features, is set to be 1800. Data from two time steps (t-1, t) are embedded into 1 layer. The spatial resolution is not reduced here considering the relative small domain. 

This data cube undergoes processing through a series of Swin Transformer V2 blocks, a variant of the Vision Transformer (ViT) that has demonstrated remarkable performance across various computer vision tasks \citep{liu2021swin}. The Swin Transformer innovatively applies self-attention mechanisms within localized windows and establishes cross-connections by shifted windows, significantly enhancing computational efficiency and effectiveness \citep{to2024architectural}. Swin Transformer V2 builds upon these strengths and utilizes a residual post normalization and a log-spaced continuous position bias technique, refining them to achieve even superior results \citep{liu2022swin}.
The architecture is organized into layers as follows:
In the first layer, the embedded data cube (H × W × C) is initially processed by two Swin Transformer blocks. Following the initial layer, a down-sampling module halves the horizontal dimensions while doubling the number of channels, transforming the data cube's dimensions to (H/2 × W/2 × 2C). The transformed data cube then passes through six Swin Transformer blocks, maintaining the dimensions of (H/2 × W/2 × 2C). The subsequent layers mirrors the structure of Layer 1 and Layer 2, maintaining the dimensions from Layer 2 and ultimately transits the data cube back to its original dimensions (H × W × C) after passing through an up-sampling module. A skip connection is established between Layer 1 and Layer 4, facilitating the concatenation along the feature channel for enriched feature representation. Lastly, a Fully Connected Layer (FCL) projects the concatenated output to generate the final predictions. This structured approach ensures efficient and effective processing, culminating in refined output predictions. More details about the choices of model architecture can be found in SI \cref{sect:si-rationale}.

\subsection{Model training}
This section describes the training details for \modelname. The model is developed on the Pytorch framework. The model is trained with 100 epochs using a batch size of 1 on each GPU. Training of the model took 100 hours on 4 Nvidia H100 GPUs. To accelerate the training process, Distributed Data Parallel (DDP) is utilized \citep{li2020pytorch}. Furthermore, to expedite data loading operations, the entire dataset is loaded into the memory of main process and distributed to worker processes during training by the main process. 
The AdamW \citep{loshchilov2019decoupled} optimizer is applied with an initial learning rate of $1\times10^{-4}$ and gradually annealed to 0 following the cosine schedule. A weight decay coefficient of 0.1 is set. To alleviate risk of over-fitting, the samples are shuffled in each epoch. The training data includes hourly fields spanning from 2018 to 2021 (35038 samples in total) and data in 2022 serves as the test dataset (8735 samples in total).

The Mean Absolute Error (MAE) loss is employed to supervise the training of the neural network. The loss function is defined as :
\begin{equation}
    \mathcal{L}=\frac{1}{C\times H\times W}\sum_{i=1}^{C}\sum_{j=1}^{H}\sum_{k=1}^{W}\Vert \mathbf{f}_{i,j,k}(\mathbf{c}_{t+1};\mathbf{c},\mathbf{w},\mathbf{e},\mathbf{b}|_{t-1,t})-\mathbf{y}_{i,j,k}\Vert^1
\end{equation}
where $C,H,W$ are the number of channels and grids in latitude and longitude, $i,j,k$ are the indices for the channels, latitude and longitude. $\mathbf{c}$, $\mathbf{w}$, $\mathbf{e}$ and $\mathbf{b}$ represent concentration, weather conditions, emissions and boundaries, respectively. $\mathbf{f}$ is the \modelname forward function. It takes the inputs at two time steps and then produces the concentration forecasts at next time step. $\mathbf{y}$ means the LOTOS-EUROS simulated concentrations, also the training target.

\subsection{Inference strategy}
The inference approach of \modelname employs an auto-regressive strategy, as illustrated in \cref{fig:architecture} (b). Initially, \modelname uses input data from the initial fields at two time steps ($t-1$ and $t$) to generate for the subsequent time point ($t+1$). Next, applying known emissions, meteorology, boundary conditions, and the forecasted concentration at $t+1$, the model predicts the concentration at $t+2$. This process enables cyclic forecasting by iteratively applying the same strategy. Each iteration takes \qty{80}{ms} on a Nvidia 4080 GPU (68.5 times faster than LOTOS-EUROS), which means it only takes less than \qty{10}{\minute} to produce 5-day hourly forecasts. In this paper, all experiments for evaluation start at 00:00 each day in 2022 and last for 5 days. The corresponding metrics are calculated on these series of forecasts and the metrics used in this research are listed in SI \cref{sect:metrics}

\section{Results}
\label{sect:results}
This section showcases the comprehensive performance of \modelname. It illustrates both the spatial and vertical error distributions to highlight the accuracy of the 3D forecasting outcomes. In addition, the trend of errors over extended forecast periods reflects \modelname's consistent reliability. To further exemplify its precision on a finer scale, example of forecasts over certain cities has been selected.

\subsection{Spatial evaluation of forecasts}
\begin{figure}
    \centering
    \includegraphics[width=1\linewidth]{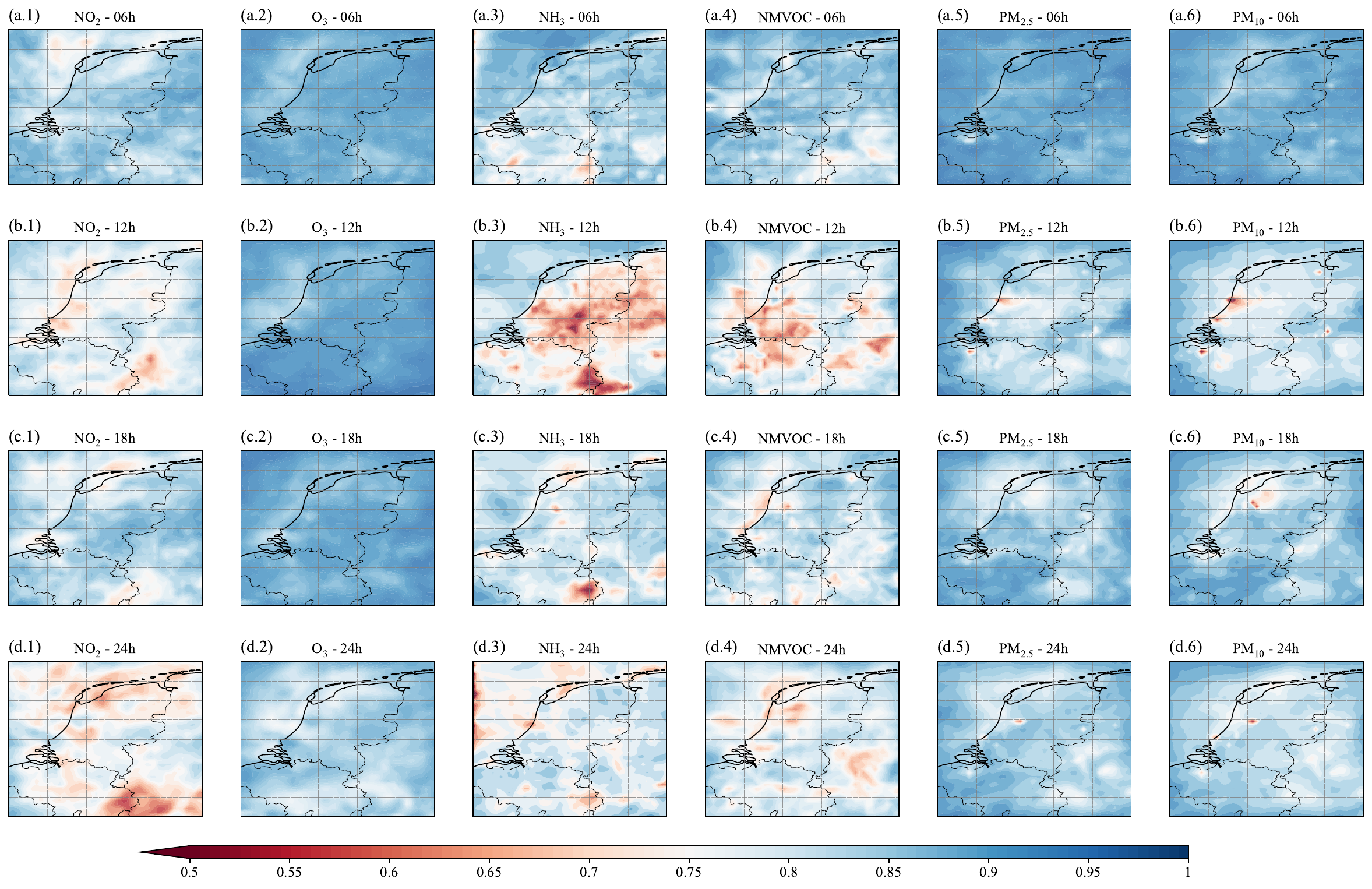}
    \caption{Spatial distribution of correlation coefficient (R) from \modelname prediction and LOTOS-EUROS simulations for variables on ground level with the lead time of 6, 12, 18, 24 hours. All the predictions start from 00:00 at each day on 2022.}
    \label{fig:spatial_r}
\end{figure}

To evaluate the overall performance of \modelname, forecasting experiments were conducted starting at 00:00 each day throughout 2022. \Cref{fig:spatial_r} illustrates the spatial distribution of the correlation coefficient (R) derived from these forecast series. Forecasts with lead times of 6, 12, 18, and 24 hours were selected to highlight trends in error and variations between day and night. The analysis includes major pollutants such as nitrogen dioxide (NO$_2$), ozone (O$_3$), ammonia (NH$_3$), non-methane volatile organic compounds (NMVOC), particulate matter less than \qty{2.5}{\micro\meter} and \qty{10}{\micro\meter} (PM$_{2.5}$ and PM$_{10}$).

The most outstanding performance is observed in the O$_3$ forecasts. As depicted in the second column of the figures, the correlation coefficient (R) for O$_3$ remains above 0.9 across most of the domain, even after 24 hours. Ozone, a secondary pollutant, is formed through complex chemical reactions involving precursor pollutants—primarily nitrogen oxides (NO$_x$) and VOCs—in the presence of sunlight. Meteorological factors, such as ambient temperature and humidity, also influence its formation. The accurate forecasting of O$_3$ indicates the superior performance of \modelname.

Forecasts for PM$_{2.5}$ and PM$_{10}$ also demonstrate strong performance, with correlation coefficients (R) exceeding 0.85 across most domains. However, some regions exhibit noticeably lower values, particularly in (c.5) and (c.6). These anomalies are primarily linked to irregular fire emissions, which vary unpredictably in both time and location. Such events lead to rapid and extreme increases in pollutant concentrations over short periods, significantly degrading evaluation metrics. This effect is not limited to PM$_{2.5}$ and PM$_{10}$ but is also observed in other species, such as NH$_3$ and NMVOC. A more detailed discussion of this phenomenon is provided in Supplementary \cref{sect:outlier}.

The performance of NO$_2$ forecasts varies across different hours. High correlation coefficients (R) are observed at 6 and 18 hours, while lower R values occur at 12 and 24 hours. The primary anthropogenic sources of NO$_2$ include the combustion of coal, oil, and gas in power plants, industrial facilities, and vehicles. As illustrated in \cref{fig:emis_day}, NO$_x$ emission trends reveal two distinct peaks around 08:00 and 16:00. \modelname effectively adapts to these emission surges, achieving strong performance during these periods. At night, in the absence of photochemical reactions, NO$_2$ enters an accumulative phase. During this time, its variations are governed by reactions with O$_3$, as well as dispersion and mixing processes. The higher R values observed at night are attributed to an underestimation of concentrations, as evidenced in \cref{fig:spatial_diff} (a.4).

The performance of NH$_3$ forecasts exhibits significant variation across different hours. At 12:00, the correlation coefficient (R) distribution reaches its lowest point. This coincides with the combined effects of peak emissions (as shown in \cref{fig:emis_day}) and enhanced dispersion driven by elevated temperatures, posing a challenge for the model to accurately capture both processes simultaneously. At other times, R remains relatively high and stable. Additionally, a notable shift in the differences between \modelname and its low-emission variant is observed, as depicted in \cref{fig:spatial_diff} (c.4). During this period, emissions are low, and a lower inversion layer restricts dispersion, leading to NH$_3$ accumulation. This suggests that while \modelname successfully captures the increasing trend of NH$_3$ at night, it may overestimate this effect.

The forecast for NMVOCs displays a trend comparable to that of NH$_3$. At 12:00, NMVOC emissions peak, contributing to elevated NH$_3$ levels. However, the accelerated rate of photochemical reactions consumes NH$_3$, while a higher inversion layer—driven by increased temperature and wind advection—adds further complexity to the dynamics. This interplay of factors challenges the model's ability to accurately capture the relationship, leading to diminished performance at noon. In contrast, at other times (06:00, 18:00, and 24:00), the model exhibits more consistent and stable performance.

\subsection{Vertical evaluation of forecasts}
\begin{figure}
    \centering
    \includegraphics[width=1\linewidth]{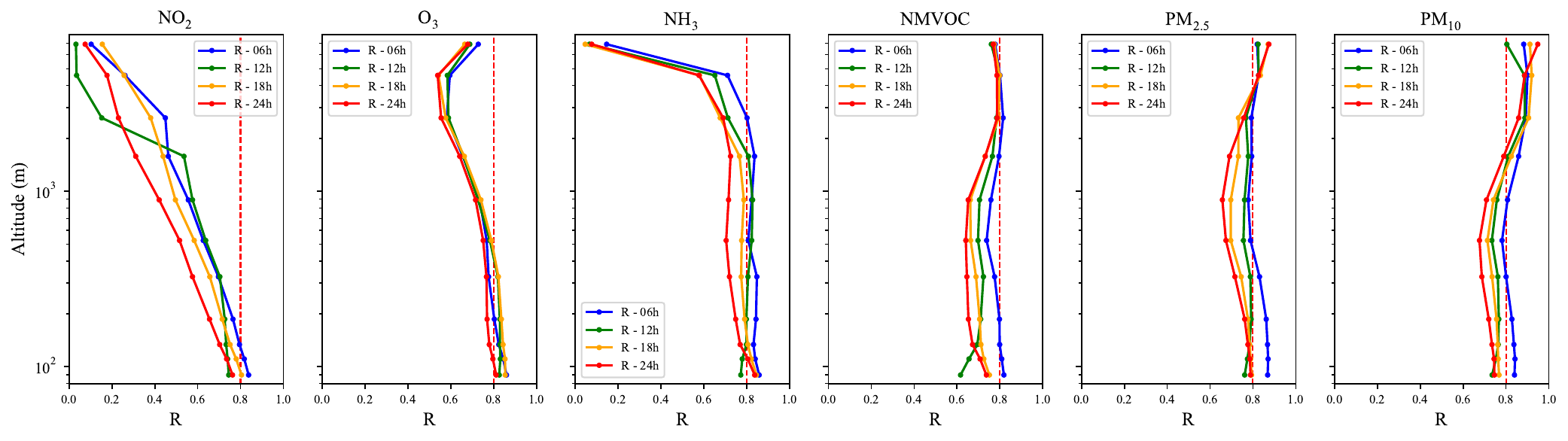}
    \caption{Vertical distribution of correlation coefficient (R) between \modelname forecast and LOTOS-EUROS with the lead time of 6, 12, 18, 24 hours. All the forecasts start from 00:00 at each day on 2022.}
    \label{fig:vertical_r}
\end{figure}

\modelname is a 3D forecast model. The forecasts verified by far are solely on the ground level. It is essential to assess the error distribution in the vertical dimension as well. \Cref{fig:vertical_r} illustrates the vertical distribution of the correlation coefficient (R) across different lead times (6, 12, 18, and 24 hours) for six major pollutants. Additional metrics, including RMSE, NMB, and NME, are presented in Supplementary \cref{fig:vertical_rmse,fig:vertical_nmb,fig:vertical_nme}. Generally, the error distribution remains consistent vertically and aligns with the concentration distribution, as shown in \cref{fig:vertical_rmse}.
The vertical R values for NMVOC, PM$_{2.5}$, and PM$_{10}$ are relatively high and stable, though they gradually decline with increasing lead time. For NH$_3$, R remains as high as 0.8 and stable across nearly all layers, except at the top layer, where it drops to near 0. Similar patterns of error variation are observed in NME and NMB. This behavior is attributed to extremely low concentration values in the upper layers, which can easily skew these metrics.
For O$_3$, R remains stable across different lead times, with an increase observed in the upper layers due to elevated concentrations near the stratosphere. Its superior performance is further evident in the low errors and biases reflected in NME, NMB, and RMSE. In contrast, NO$_2$ exhibits a continuous decrease in R with altitude, corresponding to a rapid reduction in concentration due to photochemical reactions. This diminished performance highlights \modelname's limitations in capturing this complex reaction accurately.

\subsection{Long-term evaluation of forecasts}
\begin{figure}
    \centering
    \includegraphics[width=0.6\linewidth]{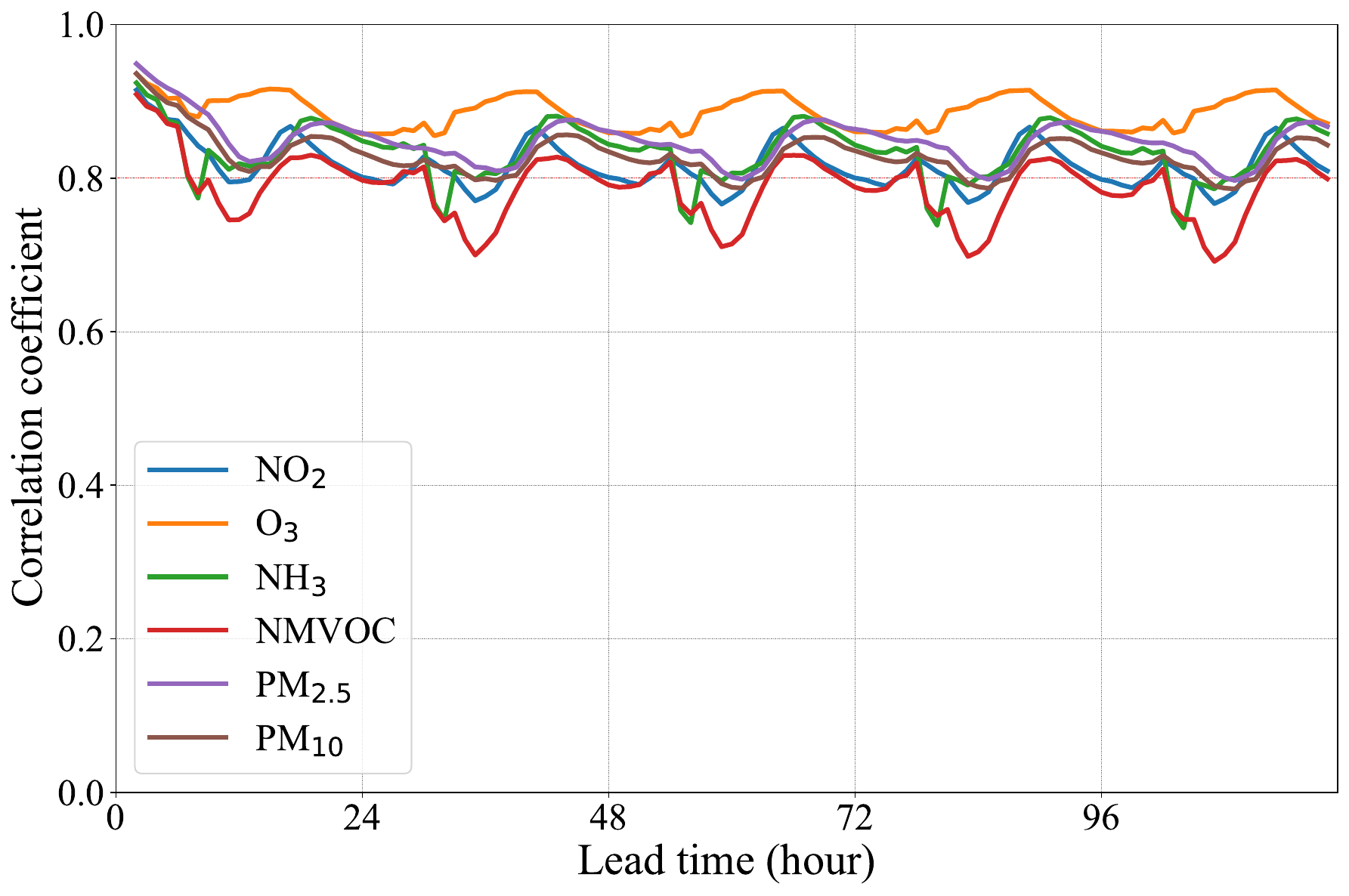}
    \caption{Trend of correlation coefficient (R) for the hourly forecasts made by \modelname. The lead time starts from 1 hour to 120 hour. The shown data is on ground level and all the forecasts start from 00:00 at each day on 2022.}
    \label{fig:sequence_r}
\end{figure}

In this section, we further evaluate the forecasting performance of \modelname over an extended lead time, with a maximum duration set to 5 days (120 hours). In practical applications, forecasts exceeding this time may be biased due to inaccuracies in weather and boundary condition predictions. \Cref{fig:sequence_r} presents the time series of the correlation coefficient (R) for six major pollutants at an hourly resolution. The results reveal a fluctuating yet generally stable performance across all pollutants. Additional metrics, including NMB, NME, and RMSE, are provided in Supplementary \cref{fig:sequence_nmb,fig:sequence_nme,fig:sequence_rmse}.

O$_3$ achieves the highest R value, consistently exceeding 0.85 across all lead times, highlighting the exceptional accuracy of \modelname in forecasting O$_3$ concentrations. Particulate matter (PM$_{2.5}$ and PM$_{10}$) exhibits similar variation patterns, with PM$_{2.5}$ slightly outperforming PM$_{10}$. The lowest R for both occurs around 14:00, coinciding with their lowest concentrations due to enhanced advection and diffusion driven by elevated temperatures, as illustrated in \cref{fig:conc_trend}. \modelname demonstrates some limitations in capturing this phenomenon effectively.

For NO$_2$, the lowest R is observed at 12:00, aligning with its minimum concentration. Similarly, NH$_3$ reaches its lowest R at 06:00, corresponding to the morning peak in NH$_3$ levels. NMVOC exhibit the most significant fluctuations, with R dropping sharply to 0.7 at 12:00 before recovering to approximately 0.8. This pattern mirrors the daily concentration trend (see \cref{fig:conc_trend}), where daytime reductions in NMVOC are attributed to photochemical reactions and a lifted inversion layer. The interplay of these factors contributes to \modelname's less consistent performance during daylight hours.

Overall, \modelname delivers robust forecasting capabilities across the evaluated pollutants, though its performance varies with diurnal cycles and specific atmospheric processes.

\begin{figure}
    \centering
    \includegraphics[width=1\linewidth]{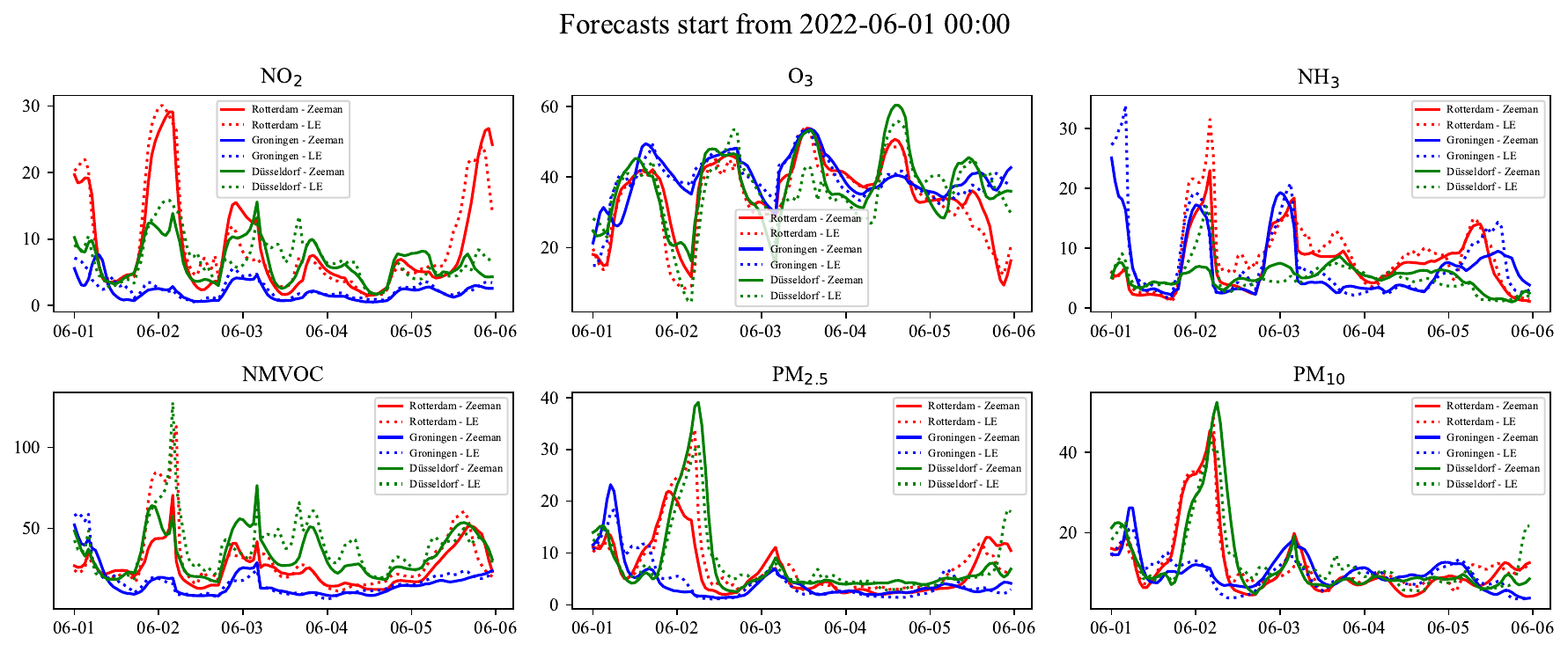}
    \caption{Time series of 5 day forecast on Rotterdam (red), Groningen (blue) and Düsseldorf (green). Dash line is LOTOS-EUROS simulations and solid line is \modelname forecast. The forecast starts from 00:00, 1st June, 2022.}
    \label{fig:point_trend}
\end{figure}

In addition, three cities (Rotterdam, Groningen, and Düsseldorf) are selected to demonstrate the forecasts of \modelname compared to LOTOS-EUROS. These cities are chosen based on their geographic locations, and their positions are plotted in \cref{fig:city_map}. \Cref{fig:point_trend} shows the forecasts of concentrations for different pollutants in these cities. The forecasts start from 2022-06-01 00:00 and last for 5 days. The solid lines represent predictions from \modelname, while the dashed lines are from LOTOS-EUROS. Additional forecasts starting on 1st March, 1st September, and 1st December are also generated and provided in the Supplementary Materials \cref{fig:point03,fig:point09,fig:point12}.

From these forecasts, a high level of agreement between \modelname and LOTOS-EUROS is evident. \modelname's forecasts of NO$_2$, O$_3$, PM$_{2.5}$ and PM$_{10}$ accurately reproduce the temporal variations and capture pollutant spikes at specific moments. For example, around midnight on 2nd June, a sharp increase in NO$_2$, PM$_{2.5}$ and PM$_{10}$ concentrations was observed. \modelname correctly predicts both the timing and magnitude of these peaks. During the normal hours when no outbreak occurs, the stable state is also reproduced. However, some discrepancies exist. For instance, while \modelname captures the peak in NMVOC concentrations around midnight on 2nd June, it shows a noticeable underestimation of the values. Overall, the results highlight the strengths of \modelname in forecasting pollutant concentrations while also identifying areas for further improvement. 

\section{Conclusion and discussions}
\label{sect:conclusion}
\subsection{Conclusion}
In this paper, we proposed a deep learning based regional atmospheric chemistry model - \modelname. It facilitates the Swin Transformer architecture to capture interplay between concentration, emission and meteorological fields. Boundary conditions are also introduced to compensate for the outer pollution transport. Hourly simulations are made using an auto-regressive strategy. Performance of \modelname simulations is evaluated. From the spatial view, high performance can be noticed in O$_3$, PM simulations with correlation coefficient exceeding 0.85 over the majority of domain. For more active species, NO$_2$, NH$_3$ and NMVOC, an increase in certain duration in the daytime can be observed, which is associated with the diurnal variation of emission and meteorology. In the vertical dimension, errors are consistent with the concentration distribution. High correlation (around 0.8) in different layers can be found in most species. Long term simulation with 5 days are presented. A stable performance for the six pollutants can be noticed. A periodic pattern is found with respect to the different time of day. Forecast series on three cities further conform the performance of \modelname. \modelname can reproduce either the stable or outbreak stage of air pollutants. 

\subsection{Limitations and future work}
While the current model represents a step forward, there are several areas where future improvements could be made. Firstly, although a range of influential factors affecting air pollutant forecasting has been considered, additional data could further enhance the model’s performance. For instance, incorporating the vertical diffusion coefficient and vertical air mass fluxes could improve the model’s ability to capture vertical transport processes more accurately. Additionally, the dataset contains limited instances of extreme weather-emission combinations, which may constrain the model’s capability to handle complex conditions effectively.

Secondly, the absence of bidirectional chemistry-meteorology coupling represents a limitation of the current model, as it restricts the ability to capture the complex, two-way interactions between atmospheric chemistry and meteorology. These interactions, such as aerosol effects on radiation budgets (e.g., direct and indirect radiative forcing) as well as their feedback on chemical transport and reactions, are important for accurately simulating real-world atmospheric dynamics. While reproducing this interaction would require a fully coupled model, which is computationally expensive and demands greater GPU resources for convergence. In present, Zeeman can act as an offline coupler to meteorological models, leveraging external meteorological fields as inputs. To enhance its utility for air quality forecasting, we envision interfacing Zeeman with existing weather models (e.g., WRF, Pangu) to incorporate dynamic meteorological outputs while maintaining a modular structure.

Thirdly, Zeeman is designed to serve as a fast, surrogate model that approximates LOTOS-EUROS’s behaviour while significantly reducing computational demands, and inaccuracies inherent in the LOTOS-EUROS may be carried over. To address this, assimilating observational data could refine \modelname and lead to more precise forecasts. Zeeman can be applied to perform efficient data assimilation, leveraging its computational speed to iteratively optimize emission estimates. With accurate emission estimates, ZEEMAN can facilitate efficient optimization of pollutant concentrations, enabling rapid and reliable predictions of atmospheric composition.

\section*{Open Research}
Zeeman was established on PyTorch, a Python based library for deep learning. In building and optimizing the backbones, Swin transformer is used and available at \url{https://github.com/microsoft/Swin-Transformer}. The user manual and inference code of \modelname are public at a GitHub repository (\url{https://github.com/xxcvvv/Zeeman}). \modelname model and dataset samples are archived on zenodo \citep{pang2025model} for users to run the test runs.

\section*{Acknowledgments}
This study was supported by the National Key Research and Development Program of China [grant number 2024YFE0113700] and the National Natural Science Foundation of China [grant number 42475150].

\bibliographystyle{unsrt}  
\bibliography{my_lib}

\newpage
\appendix

\large{\textbf{Appendix}}

\section{Names of emissions}
\begin{table}[H]
    \centering
    \caption{Full name of the emissions}
    \label{tab:names}
    \begin{tabular}[c]{llll}
        \toprule
        Emission  &  Full name & Emission  &  Full name \\
        \midrule
        no2 & Nitrogen Dioxide & ch4 & Methane \\
        no & Nitric Oxide & nh3 & Ammonia \\
        co & Carbon Monoxide & iso & Isoprene \\
        form & Formaldehyde & terp & Terpenes \\
        ald & Aldehydes (general) & ec\_f & Elemental Carbon (Fine) \\
        par & Peroxyacyl Nitrates & ec\_c & Elemental Carbon (Coarse) \\
        ole & Olefins (Alkenes) & pom\_f & Particulate Organic Matter (Fine) \\
        eth & Ethane & pom\_c & Particulate Organic Matter (Coarse) \\
        tol & Toluene & ppm\_f & Potassium Permanganate (Fine) \\
        xyl & Xylene & ppm\_c & Potassium Permanganate (Coarse) \\
        so4a\_f & Sulfate Aerosol (Fine) & na\_f \& na\_ff & Sodium (Fine mode) \\
        so2 & Sulfur Dioxide & na\_c \& na\_cc \& na\_ccc & Sodium (Coarse mode) \\
        \bottomrule
    \end{tabular}
\end{table}

\section{Evaluation metrics}
\label{sect:metrics}
The performance of the \modelname is evaluated using the following metrics: the correlation coefficient (R), the root mean square error (RMSE), and the normalized mean error (NME). These metrics are calculated as follows:
\begin{equation}
    R=
    \frac{\sum_{i=1}^{H}\sum_{j=1}^{W}(\mathbf{y}_{i,j}-\overline{\mathbf{y}}_{i,j})(\mathbf{c}_{i,j}-\overline{\mathbf{c}}_{i,j})}
    {\sqrt{\sum_{i=1}^{H}\sum_{j=1}^{W}(\mathbf{y}_{i,j}-\overline{\mathbf{y}}_{i,j})^{2}\sum_{i=1}^{H}\sum_{j=1}^{W}(\mathbf{c}_{i,j}-\overline{\mathbf{c}}_{i,j})^{2}}}
\end{equation}
\begin{equation}
    RMSE=
    \sqrt{\frac{\sum_{i=1}^{H}\sum_{j=1}^{W}(\mathbf{y}_{i,j}-\mathbf{c}_{i,j})^{2}}{H\times W}}
\end{equation}
\begin{equation}
    NME=
    \frac{\sum_{i=1}^{H}\sum_{j=1}^{W}\left|\mathbf{y}_{i,j}-\mathbf{c}_{i,j}\right|}
    {\sum_{i=1}^{H}\sum_{j=1}^{W}\mathbf{y}_{i,j}}
\end{equation}
\begin{equation}
    NMB=
    \frac{\sum_{i=1}^{H}\sum_{j=1}^{W}(\mathbf{y}_{i,j}-\mathbf{c}_{i,j})}
    {\sum_{i=1}^{H}\sum_{j=1}^{W}\mathbf{y}_{i,j}}
\end{equation}
$i$ and $j$ represent the indices of latitude and longitude. $\mathbf{y}_{i,j}$ and $\mathbf{c}_{i,j}$ are the target and predicted values at a specific latitude and longitude.

\section{Dataset overview}
\label{sect:si-dataset}
\subsection{Daily emission trends}
\begin{figure}[ht]
    \centering
    \includegraphics[width=0.8\textwidth]{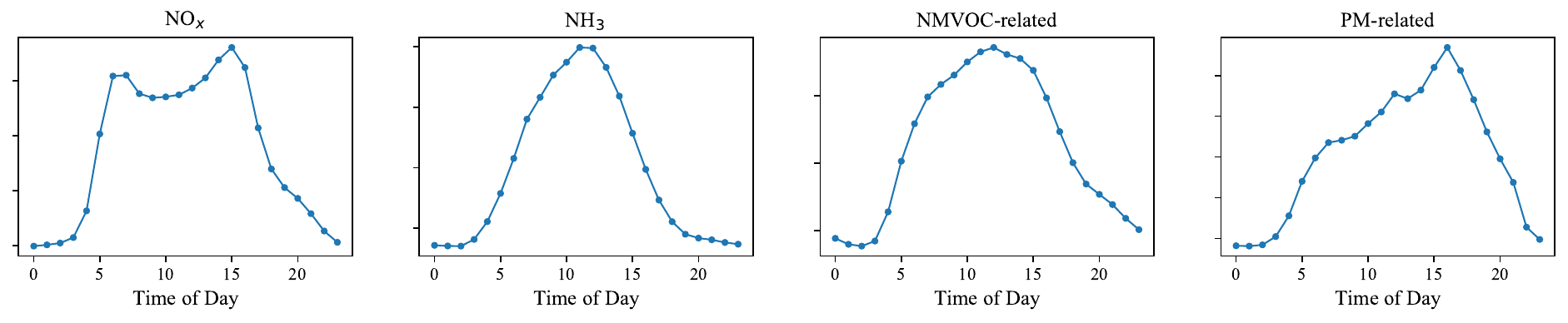}
    \caption{Trend of emissions in different time of day at 2022. }
    \label{fig:emis_day}
\end{figure}

\Cref{fig:emis_day} shows the trend of emissions in different time of a day. They are averaged from hourly emissions in 2022. In detail, the composition of the emissions is:
\begin{itemize}
    \item NO$_{x}$ = no + no2
    \item NH$_3$ = nh3
    \item NMVOC-related = form + ald + par + ole + eth + tol + xyl + so4a\_f + iso + terp
    \item PM-related = ec\_f + ec\_c + pom\_f + pom\_c + ppm\_f + ppm\_c + na\_ff + na\_f + na\_ccc + na\_cc + na\_c
\end{itemize}
\begin{figure}[H]
    \centering
    \includegraphics[width=0.7\textwidth]{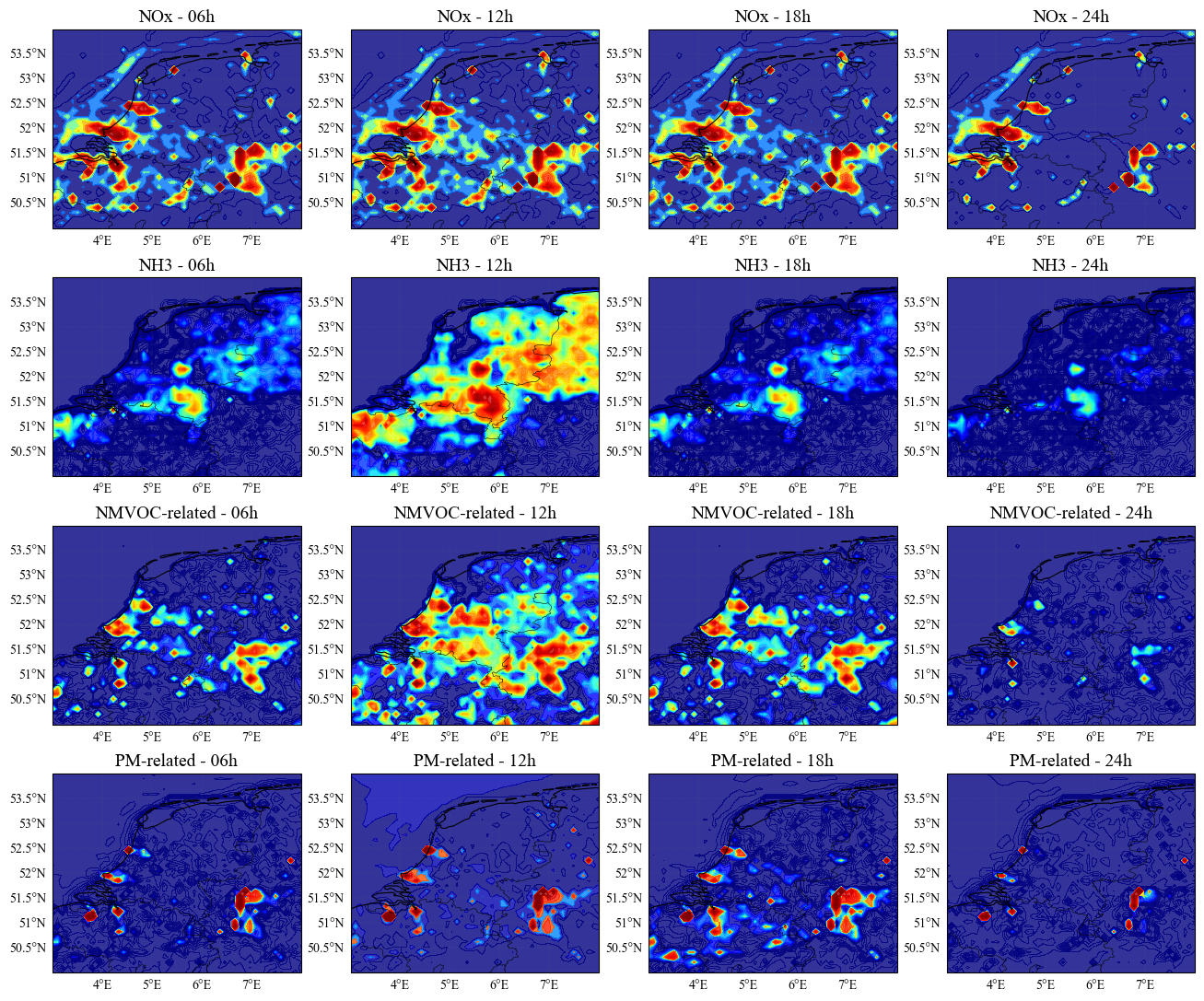}
    \caption{Spatial distribution of averaged emissions in different time of day at 2022. }
    \label{fig:emis_spatial}
\end{figure}

\subsection{Yearly emission trends}
\begin{figure}[ht]
    \centering
    \includegraphics[width=1\linewidth]{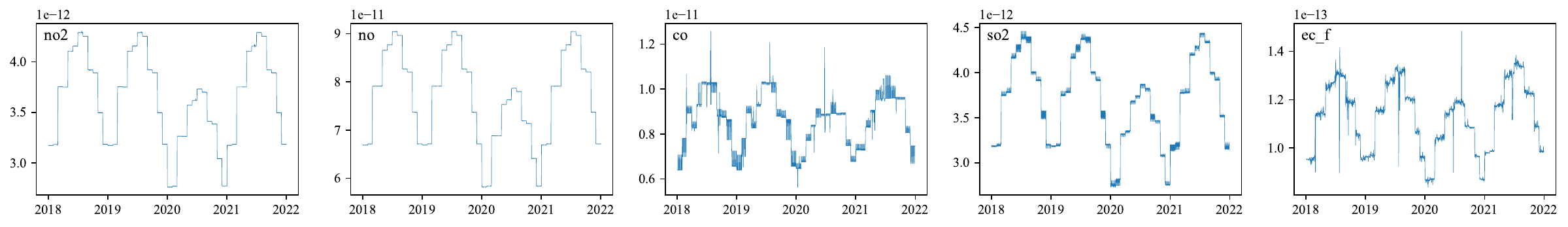}
    \caption{Time series of emission in Utrecht from 2018 to 2021.}
    \label{fig:emis_series}
\end{figure}

During epidemics, there was a noticeable reduction in emission. We added a time series of emission in Utrecht across the time span as shown below \cref{fig:emis_series}. There is a clear reduction in emission in 2020 when the pandemic starts.

\subsection{Trend of concentrations}
\begin{figure}[H]
    \centering
    \includegraphics[width=0.7\textwidth]{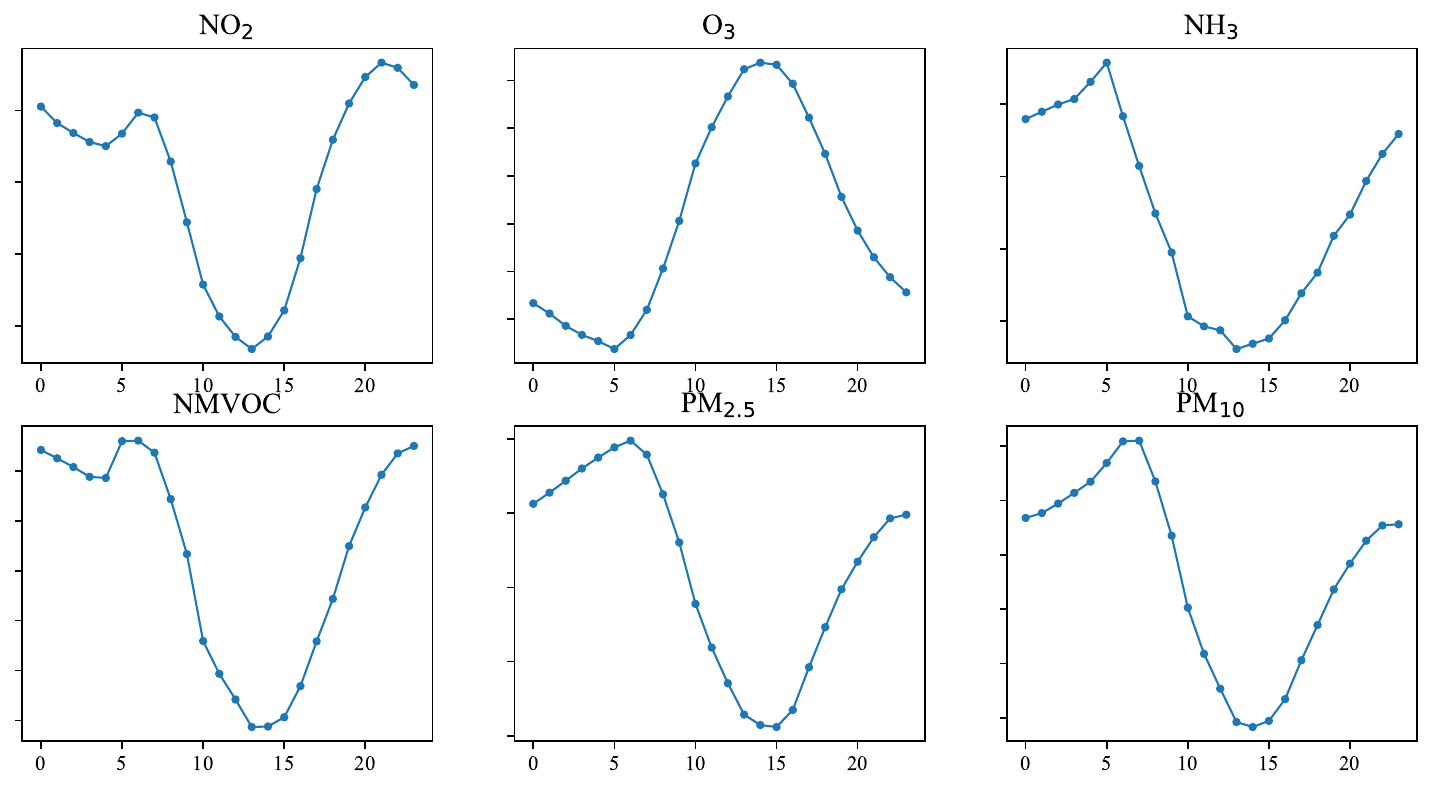}
    \caption{Trend of averaged concentrations from LOTOS-EUROS simulations in different time of day at 2022. }
    \label{fig:conc_trend}
\end{figure}

\subsection{Weather conditions}
\begin{figure}[H]
    \centering
    \includegraphics[width=0.4\linewidth]{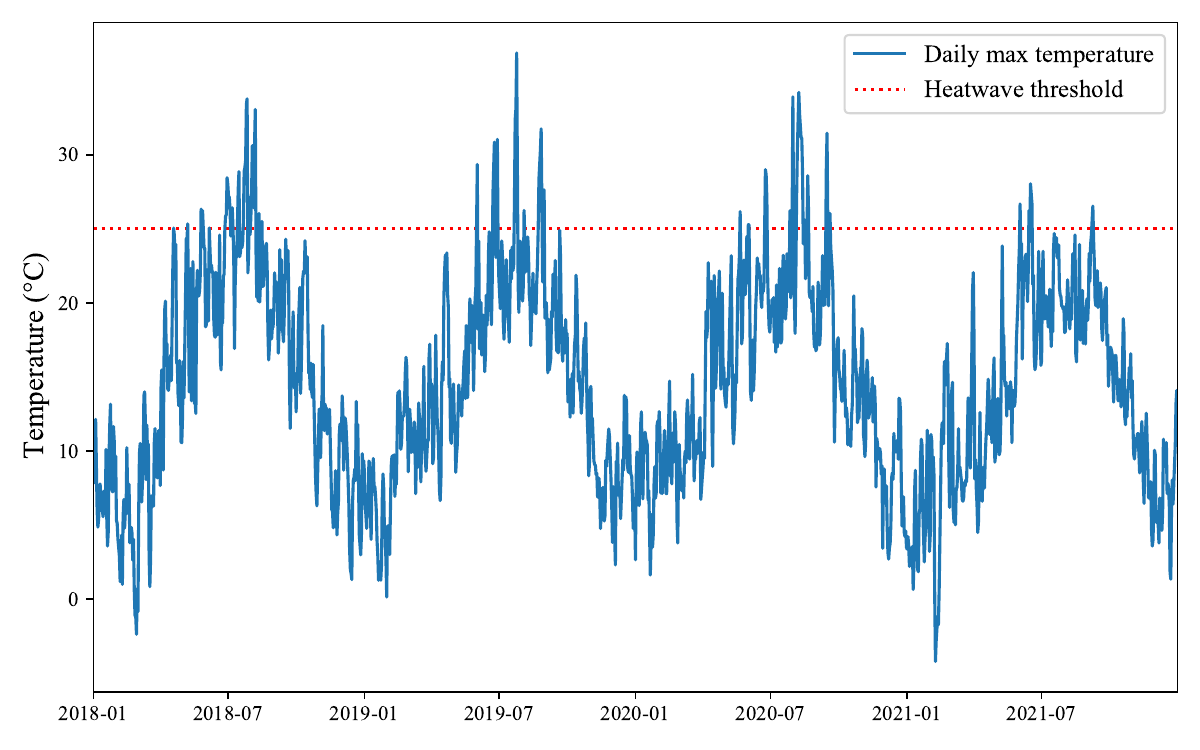}
    \caption{Time series of daily maximum temperature in Maastricht from 2018 to 2021.}
    \label{fig:heatwave}
\end{figure}

We checked the weather conditions in this time period. It covers various weather conditions. For example, heat waves are assessed in the dataset. In the Netherlands, a heatwave is defined as at least five consecutive days with a maximum temperature of 25\unit{\celsius} or above. It is a relatively low criterion for the high latitude of research area. \Cref{fig:heatwave} shows the time series of daily max temperature in Maastricht from 2018 to 2021. According to statistics, there are 88 days that max temperature exceeds 25\unit{\celsius} and 14 heatwave events detected across these four years. 

\section{Rationale behind model architecture}
\label{sect:si-rationale}
In general, owing to heavy training cost overhead, we didn't perform experiments on ablation studies. We believe that current configurations balance the cost and accuracy. The performance of swin transformer to handle high dimensional data has been proven in many AI weather prediction models \citep{bi2023accurate,chen2023fuxi}. \cite{to2024architectural} has performed an ablation study on light Pangu model and results show that 2D Transformer architecture yields a model that is more robust to training, converges faster, and produces better forecasts compared to using the 3D Transformer. We also admit there are settings that can bring higher accuracy or training efficiency. We added discussions behind the choice of architecture as shown below:

The Zeeman model, designed for complex tasks, processes a substantial input with 17 species across 11 atmospheric layers, resulting in 187 input channels. This high-dimensional input encapsulates a wide range of chemical concentrations, across different altitudes. Capturing the intricate relationships and dependencies between these channels requires a robust feature representation, which is achieved by projecting the input data into a high-dimensional feature space with a large number of feature channels.
For context, consider the deep learning-based weather prediction model Fuxi, which handles 70 input channels—representing meteorological variables and projects them into a feature space with 1536 channels \citep{chen2023fuxi}. This expansion allows Fuxi to model complex, non-linear relationships between variables, enabling accurate predictions. Similarly, in the Zeeman model, a large feature channel dimension is essential to encode the complex interplay between the input channels, such as how different species interact across atmospheric layers or how their properties vary with altitude.
A larger number of feature channels enhances the model’s capacity to learn and represent these intricate relationships, potentially improving its ability to capture subtle patterns and dependencies. For instance, in atmospheric modeling, interactions between species (e.g., chemical reactions) and their variations across layers can be highly non-linear, necessitating a high-dimensional feature dimension to avoid underfitting and ensure sufficient expressive power.
However, this approach comes with a trade-off: increasing the number of feature channels can heighten the risk of overfitting, especially if the model memorizes noise or spurious patterns in the training data rather than generalizing to unseen data. Overfitting is particularly concerning in high-dimensional models like Zeeman, where the large input space and feature channel count amplify the model’s complexity. To mitigate this, techniques such as regularization (e.g., dropout, weight decay) are applied to ensure the model generalizes well.

The layer structure follows the structure in \cite{bi2023accurate}. It is significant fewer than standard Swin transformer. This is to reduce the complexity of computation and memory.

Larger batch sizes contribute to smoother and more stable gradients during training. This stability arises because gradients are computed as an average over a larger number of samples, reducing the variance introduced by noisy or outlier data points. While it can also smooth out pollution outburst events, which is crucial to atmospheric chemistry model.  
On the other hand, the memory cost typically scales linearly with batch size for the data-related components (e.g., input tensors and activations. For a big model like Zeeman, which costs large GPU memory to train, batch size of 1 is a reasonable choice to contain the memory cost. Besides, during evaluation or inference, a batch size of 1 is often sufficient, as predictions are made one sample at a time.
    
\section{Inference speed}
\label{sect:si-speed}
\Cref{tab:comparison} lists the inference speed and hardware requirements for running Zeeman and LOTOS-EUROS. Computation speed of Zeeman (on 1 4080 gpu) is 68.5 times faster than LOTOS-EUROS (on 16 cpu cores) given the domain and resolution presented in the manuscript. Note that LOTOS-EUROS's runtime is averaged over a six-year simulation. In practice, the initialization of numerical models can be time-intensive compared to individual time steps, potentially making Zeeman’s speed over 100 times faster.

LOTOS-EUROS's computational speed is generally linear with the number of grid points. In the case of inference, Zeeman also has linear scalability thanks to the window attention mechanism in Swin transformer \citep{liu2021swin}.

\begin{table}[ht]
    \centering
    \caption{Computational demands for running LOTOS-EUROS and Zeeman.}
    \label{tab:comparison}
    \begin{tabular}[c]{cccc}
    \toprule
        Name & Model & Numbers & Notes \\
    \midrule
        Runtime (per step/hour) & 
        \makecell[c]{LOTOS-EUROS\\Zeeman} & 
        \makecell[c]{5479 ms (1$\times$) \\80 ms (68.5$\times$)} & 
        \makecell[c]{16$\times$cpu cores \\  1$\times$4080 gpu} \\
        Hardware & 
        \makecell[c]{LOTOS-EUROS\\Zeeman} & 
        \makecell[c]{16g cpu memory\\ 10g gpu memory} & 
        \makecell[c]{16$\times$cpu cores\\ 1$\times$gpu }\\
    \bottomrule
    \end{tabular}
\end{table}

\section{Necessity of boundary conditions}
\label{sect:si-boundary}
For a regional atmospheric chemistry model, boundary conditions plays an important role especially after lone time forecasting \citep{qu2024effect}. To address this concern, we have added a set of experiment without boundary inputs to show the differences. \Cref{tab:boundary} lists the Normalized Mean Bias (NMB) in four boundaries from 3h forecast made with or boundary conditions. For most of the variables, forecast with boundary exhibits lower bias. It confirms the necessity of adding boundary condition to Zeeman.

\begin{table}[ht]
    \centering
    \caption{Comparison of NMB in four boundaries from 3h forecast made with or boundary conditions.}
    \label{tab:boundary}
    \begin{tabular}[c]{cccccc}
    \toprule
        Name & Type & West & East & South & North \\
    \midrule
        NO$_2$ & \makecell[c]{with boundary \\ without boundary} & \makecell[c]{\textbf{-0.043} \\ -0.048 } & \makecell[c]{\textbf{-0.023} \\ -0.052 } & \makecell[c]{\textbf{-0.013} \\ -0.035 } & \makecell[c]{\textbf{-0.054} \\ -0.077 } \\
        O$_3$ & \makecell[c]{with boundary \\ without boundary} & \makecell[c]{0.014 \\ \textbf{0.009} } & \makecell[c]{-0.001 \\ 0.001 } & \makecell[c]{0.025 \\ 0.021} & \makecell[c]{0.019 \\ 0.020 } \\
        CO & \makecell[c]{with boundary \\ without boundary} & \makecell[c]{-0.021 \\ \textbf{-0.012} } & \makecell[c]{-0.013 \\ -0.016 } & \makecell[c]{\textbf{-0.009} \\ -0.016 } & \makecell[c]{0.014 \\ 0.017 } \\
        SO$_2$ & \makecell[c]{with boundary \\ without boundary} & \makecell[c]{\textbf{0.008} \\ -0.406 } & \makecell[c]{\textbf{-0.167} \\ -0.452 } & \makecell[c]{\textbf{-0.095} \\ -0.589 } & \makecell[c]{\textbf{-0.135} \\ -1.273 } \\
        NH$_3$ & \makecell[c]{with boundary \\ without boundary} & \makecell[c]{\textbf{-0.096} \\ -0.127 } & \makecell[c]{0.005 \\ -0.009 } & \makecell[c]{-0.090 \\ -0.091 } & \makecell[c]{\textbf{-0.016} \\ -0.246 } \\
        NH$_4$ & \makecell[c]{with boundary \\ without boundary} & \makecell[c]{\textbf{0.017} \\ 0.285 } & \makecell[c]{\textbf{0.013} \\ 0.190 } & \makecell[c]{\textbf{0.003} \\ 0.174 } & \makecell[c]{\textbf{0.033} \\ 0.418 } \\
        PAN & \makecell[c]{with boundary \\ without boundary} & \makecell[c]{\textbf{0.069} \\ -0.741 } & \makecell[c]{\textbf{-0.046} \\ -0.710 } & \makecell[c]{\textbf{-0.330} \\ -0.905 } & \makecell[c]{\textbf{-0.004} \\ -0.848 } \\
        SO$_4$ & \makecell[c]{with boundary \\ without boundary} & \makecell[c]{\textbf{-0.033} \\ 0.091 } & \makecell[c]{-0.095 \\ \textbf{-0.011} } & \makecell[c]{-0.072 \\ \textbf{0.007} } & \makecell[c]{\textbf{-0.032} \\ 0.131 } \\
        NO$_3$ & \makecell[c]{with boundary \\ without boundary} & \makecell[c]{\textbf{0.023} \\ 0.052 } & \makecell[c]{\textbf{0.055} \\ 0.067 } & \makecell[c]{-0.005 \\ 0.010 } & \makecell[c]{\textbf{0.022} \\ 0.081 } \\
        NO$_3$ & \makecell[c]{with boundary \\ without boundary} & \makecell[c]{\textbf{-0.003} \\ -0.289 } & \makecell[c]{\textbf{0.056} \\ -0.321 } & \makecell[c]{\textbf{-0.078} \\ -0.468 } & \makecell[c]{\textbf{-0.018} \\ -0.286 } \\
        EC & \makecell[c]{with boundary \\ without boundary} & \makecell[c]{\textbf{-0.008} \\ 0.193 } & \makecell[c]{\textbf{-0.059} \\ 0.103 } & \makecell[c]{-0.150 \\ \textbf{0.024} } & \makecell[c]{\textbf{-0.010} \\ 0.496 } \\
        POM & \makecell[c]{with boundary \\ without boundary} & \makecell[c]{-0.048 \\ \textbf{-0.033} } & \makecell[c]{-0.010 \\ -0.007 } & \makecell[c]{-0.071 \\ -0.066 } & \makecell[c]{\textbf{0.009} \\ 0.035 } \\
        PPM & \makecell[c]{with boundary \\ without boundary} & \makecell[c]{\textbf{-0.082} \\ -0.186 } & \makecell[c]{-0.020 \\ \textbf{-0.001} } & \makecell[c]{\textbf{-0.049} \\ -0.252 } & \makecell[c]{\textbf{0.035} \\ -0.478 } \\
        NMVOC & \makecell[c]{with boundary \\ without boundary} & \makecell[c]{\textbf{-0.032} \\ 0.008 } & \makecell[c]{-0.016 \\ -0.021 } & \makecell[c]{\textbf{-0.005} \\ -0.017 } & \makecell[c]{-0.009 \\ 0.006 } \\
        PM$_{2.5}$ & \makecell[c]{with boundary \\ without boundary} & \makecell[c]{\textbf{-0.008} \\ 0.019 } & \makecell[c]{\textbf{0.026} \\ 0.040 } & \makecell[c]{\textbf{0.001} \\ 0.014 } & \makecell[c]{\textbf{-0.006} \\ 0.028 } \\
        PM$_{10}$ & \makecell[c]{with boundary \\ without boundary} & \makecell[c]{\textbf{-0.015} \\ 0.042 } & \makecell[c]{\textbf{0.023} \\ 0.083 } & \makecell[c]{\textbf{-0.004} \\ 0.067 } & \makecell[c]{\textbf{0.009} \\ 0.058 } \\
        Sea spray & \makecell[c]{with boundary \\ without boundary} & \makecell[c]{\textbf{-0.010} \\ 0.069 } & \makecell[c]{\textbf{0.013} \\ 0.210 } & \makecell[c]{\textbf{-0.004} \\ 0.303 } & \makecell[c]{\textbf{0.002} \\ 0.046 }\\
    \bottomrule
    \end{tabular}
\end{table}

\section{Validation with observation}
\label{sect:si-validate}
\begin{figure}[ht]
    \centering
    \includegraphics[width=0.8\linewidth]{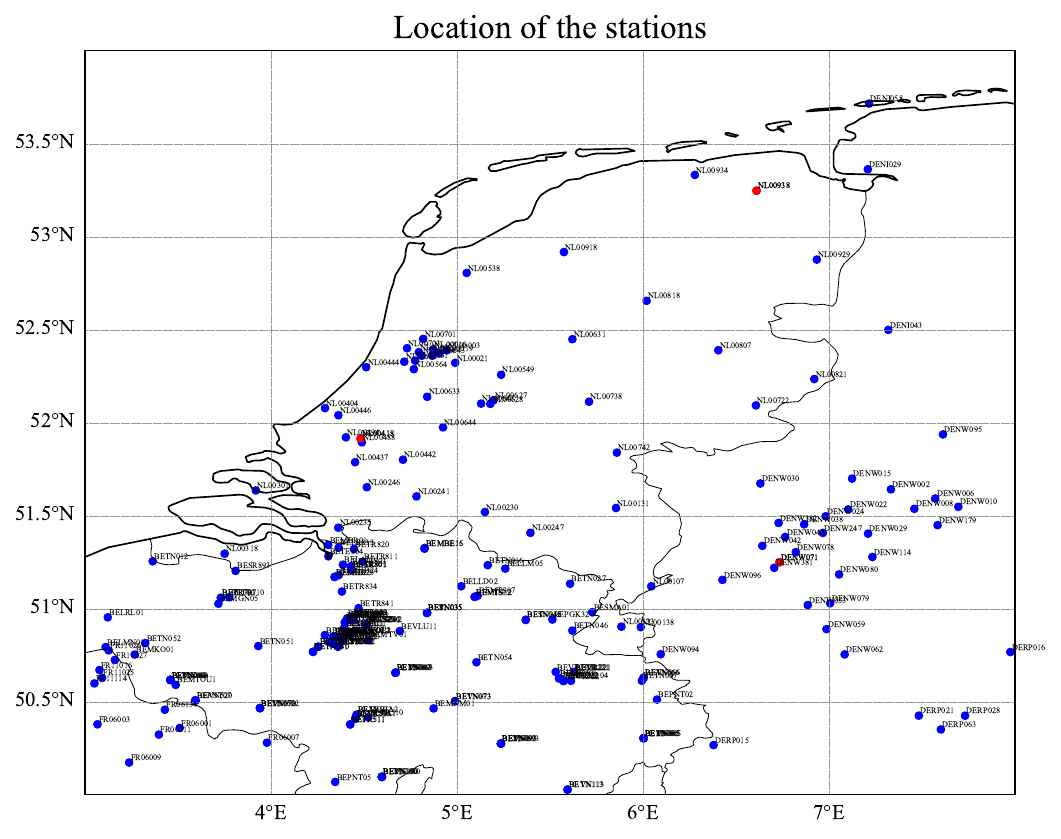}
    \caption{Spatial distribution of the ground monitor stations}
    \label{fig:spatial_obs}
\end{figure}

We validate the LOTOS-EUROS and Zeeman simulations with ground observations. \Cref{fig:spatial_obs} illustrates the spatial distribution of ground monitor stations inside the research domain. Here, stations in Rotterdam, Groningen and Düsseldorf are chosen to validate the simulations from LOTOS-EUROS and Zeeman, which is consistent with the three cities selected in the manuscript. \Cref{tab:obs_validate} shows the RMSE and NMB of four pollutants (NO$_2$, O$_3$, PM$_{2.5}$ and PM$_{10}$). These metrics are calculated from four 5-day simulation experiments (starts from 1th Mar., Jun., Sep. and Dec. 2022.). It can be found that Zeeman exhibits comparable accuracy to LOTOS-EUROS. In Rotterdam and Düsseldorf, there is slightly better performance than LOTOS-EUROS.

\begin{table}[ht]
    \centering
    \caption{Performance validation of Zeeman.}
    \label{tab:obs_validate}
    \begin{tabular}{lccccc}
        \toprule
            Places & Models & Variables & RMSE & NMB \\
        \midrule
            \multirow{4}{*}{Rotterdam} & LOTOS-EUROS & \makecell{NO$_2$\\ O$_3$\\PM$_{2.5}$\\ PM$_{10}$} & \makecell{23.01 \\ 15.07 \\ 9.28 \\ 11.84} & \makecell{-0.54 \\ -0.27 \\ 0.27 \\ 0.15} \\
             & Zeeman & \makecell{NO$_2$\\ O$_3$\\PM$_{2.5}$\\ PM$_{10}$} & \makecell{22.72 \\ 13.34 \\ 8.15 \\ 11.68} & \makecell{-0.55 \\ -0.28 \\ 0.20 \\ 0.14} \\
        \midrule
            \multirow{4}{*}{Groningen} & LOTOS-EUROS & \makecell{NO$_2$\\ O$_3$\\PM$_{2.5}$\\ PM$_{10}$} & \makecell{8.60 \\ 15.86 \\ 6.08 \\ N/A} & \makecell{-0.41 \\ -0.24 \\ 0.04 \\ N/A} \\
             & Zeeman & \makecell{NO$_2$\\ O$_3$\\PM$_{2.5}$\\ PM$_{10}$} & \makecell{8.81 \\ 14.81 \\ 6.13 \\ N/A} & \makecell{-0.51 \\ -0.21 \\ -0.04 \\ N/A} \\
        \midrule
            \multirow{4}{*}{Düsseldorf} & LOTOS-EUROS & \makecell{NO$_2$\\ O$_3$\\PM$_{2.5}$\\ PM$_{10}$} & \makecell{23.24 \\ 24.02 \\ 5.66 \\ 7.77} & \makecell{-0.53 \\ 1.50 \\ -0.07 \\ 0.07} \\
             & Zeeman & \makecell{NO$_2$\\ O$_3$\\PM$_{2.5}$\\ PM$_{10}$} & \makecell{21.43 \\ 23.50 \\ 5.81 \\ 7.39} & \makecell{-0.50 \\ 1.35 \\ -0.05 \\ 0.08} \\
        \bottomrule
    \end{tabular}
\end{table}

\section{Spatial error distribution}
\begin{figure}[ht]
    \centering
    \includegraphics[width=1\linewidth]{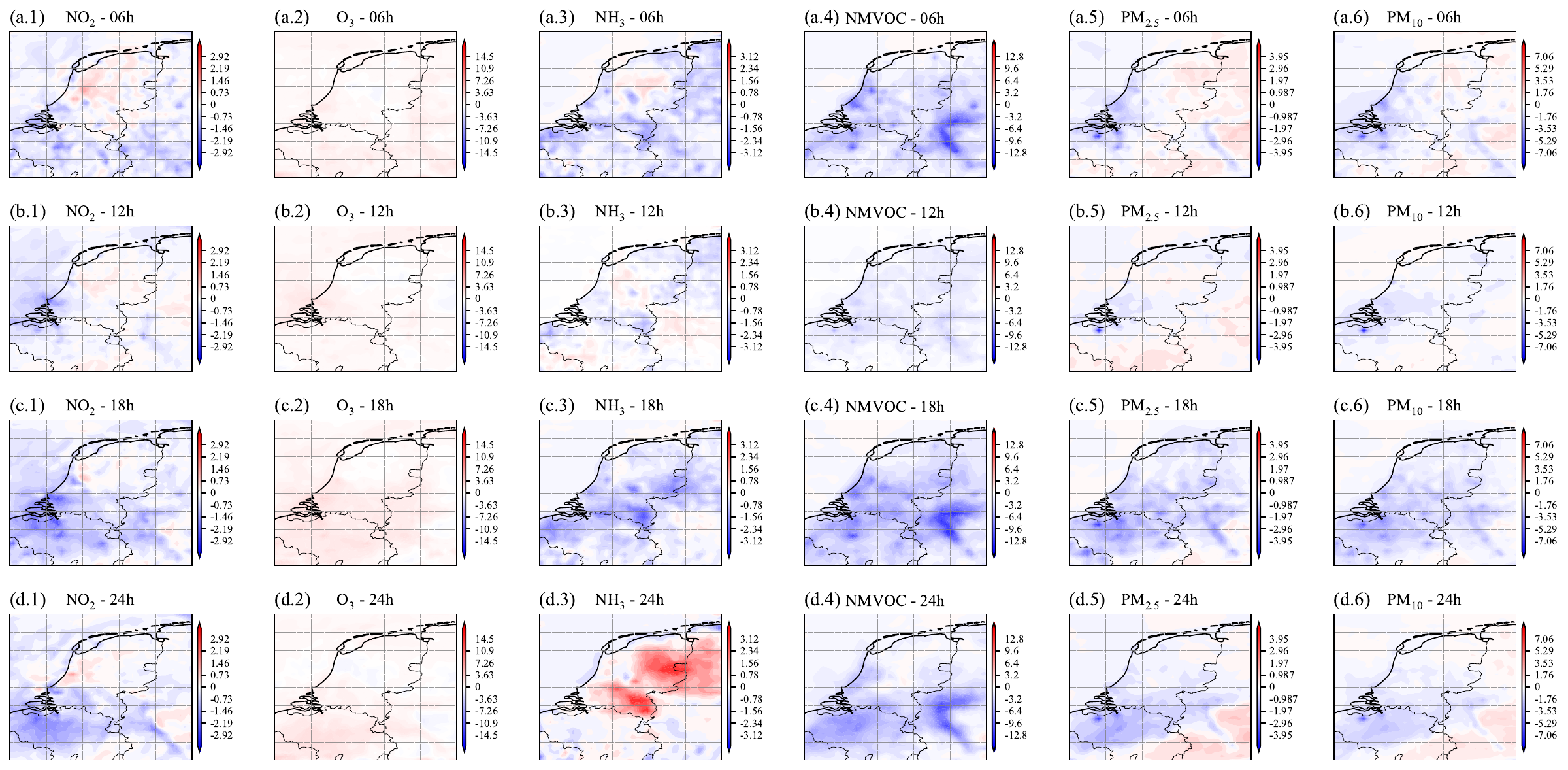}
    \caption{Spatial distribution of differences between LOTOS-EUROS simulations and forecasts made by \modelname with lead time of 6, 12, 18, 24 hours. The shown data is on ground level and all the forecasts start from 00:00 at each day on 2022.}
    \label{fig:spatial_diff}
\end{figure}
\begin{figure}[ht]
    \centering
    \includegraphics[width=1\linewidth]{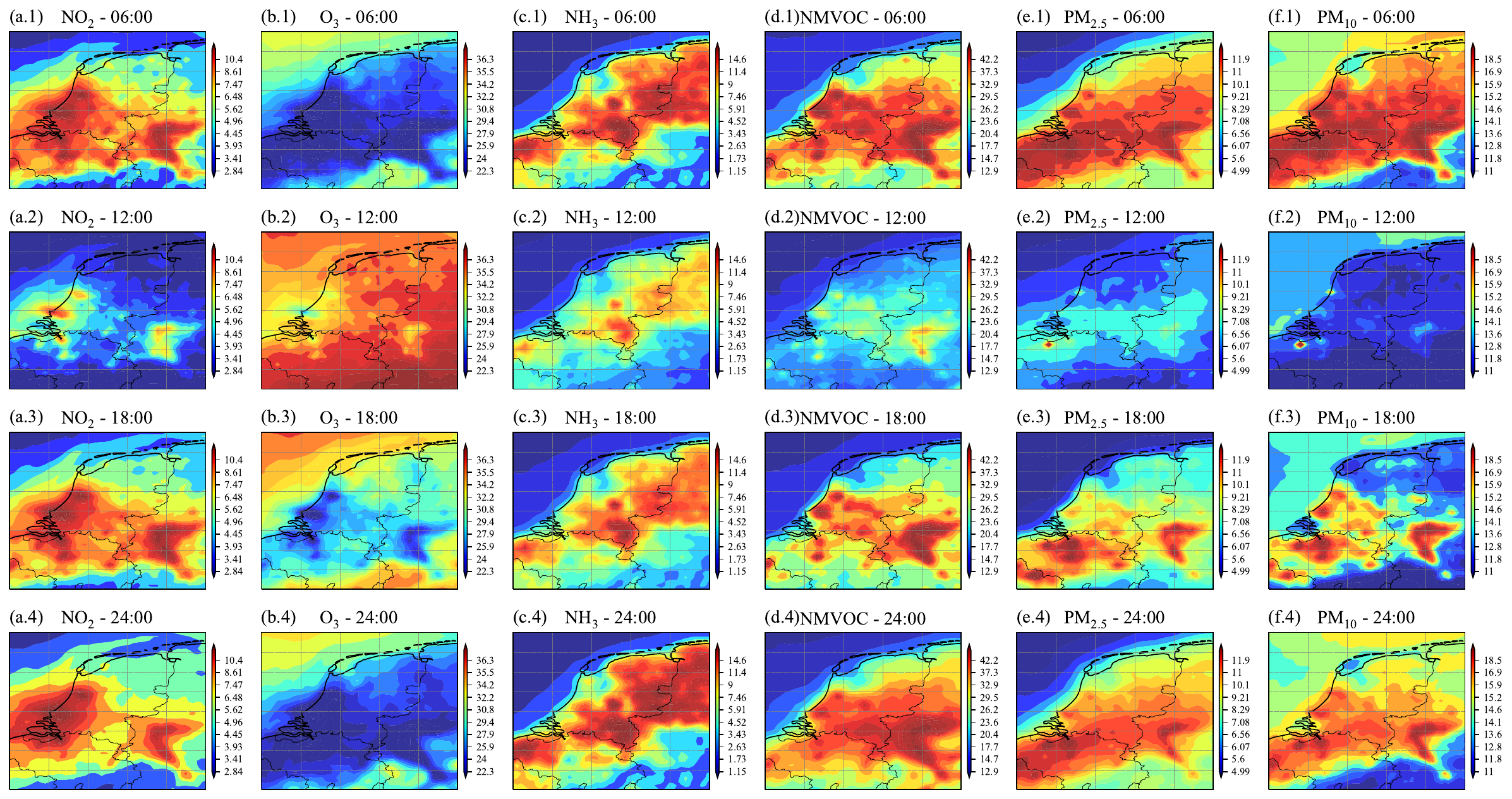}
    \caption{Spatial distribution of averaged concentrations from \modelname simulations at 06:00, 12:00, 18:00, 24:00 of day. The shown data is on ground level and averaged on 2022.}
    \label{fig:spatial_mean}
\end{figure}

\section{Vertical error distribution}
Vertical error distributions of \modelname forecasts including RMSE, NMB and NME are shown here.
\begin{figure}[ht]
    \centering
    \includegraphics[width=1\linewidth]{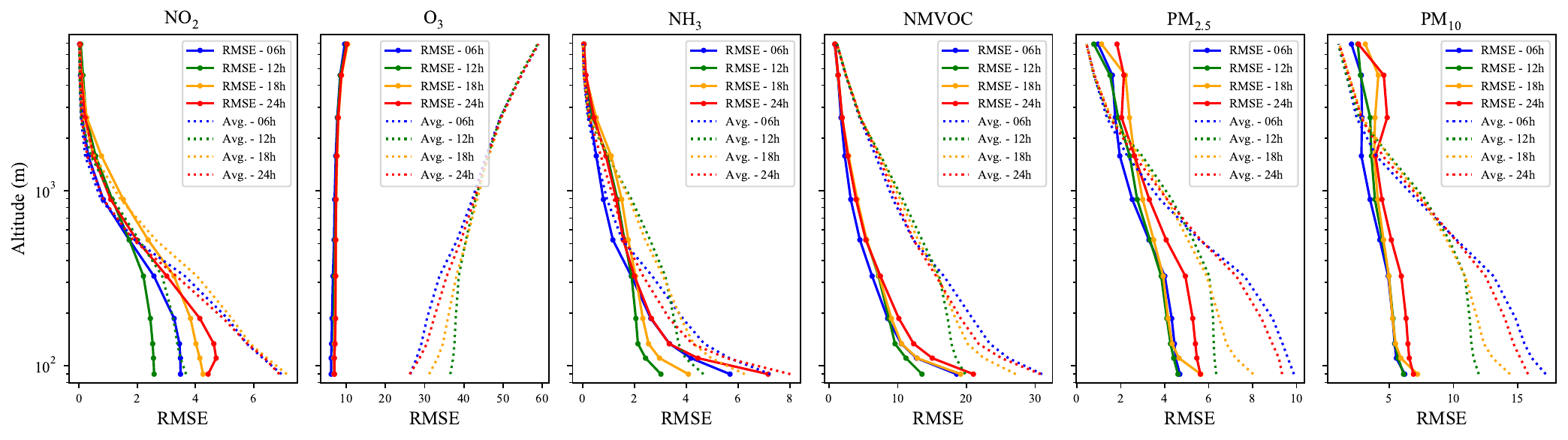}
    \caption{Vertical distribution of RMSE between \modelname forecast and LOTOS-EUROS for all the variables with the lead time of 6, 12, 18, 24 hours. All the predictions start from 00:00 at each day on 2022.}
    \label{fig:vertical_rmse}
\end{figure}
\begin{figure}[ht]
    \centering
    \includegraphics[width=1\linewidth]{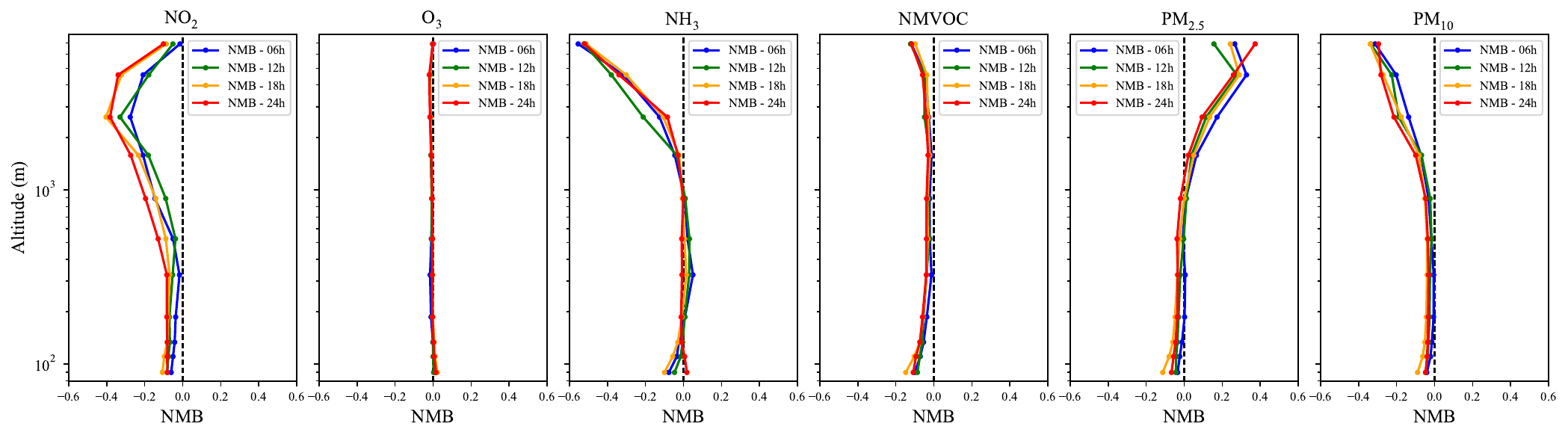}
    \caption{Vertical distribution of NMB between \modelname forecast and LOTOS-EUROS for all the variables with the lead time of 6, 12, 18, 24 hours. All the predictions start from 00:00 at each day on 2022.}
    \label{fig:vertical_nmb}
\end{figure}
\begin{figure}[ht]
    \centering
    \includegraphics[width=1\linewidth]{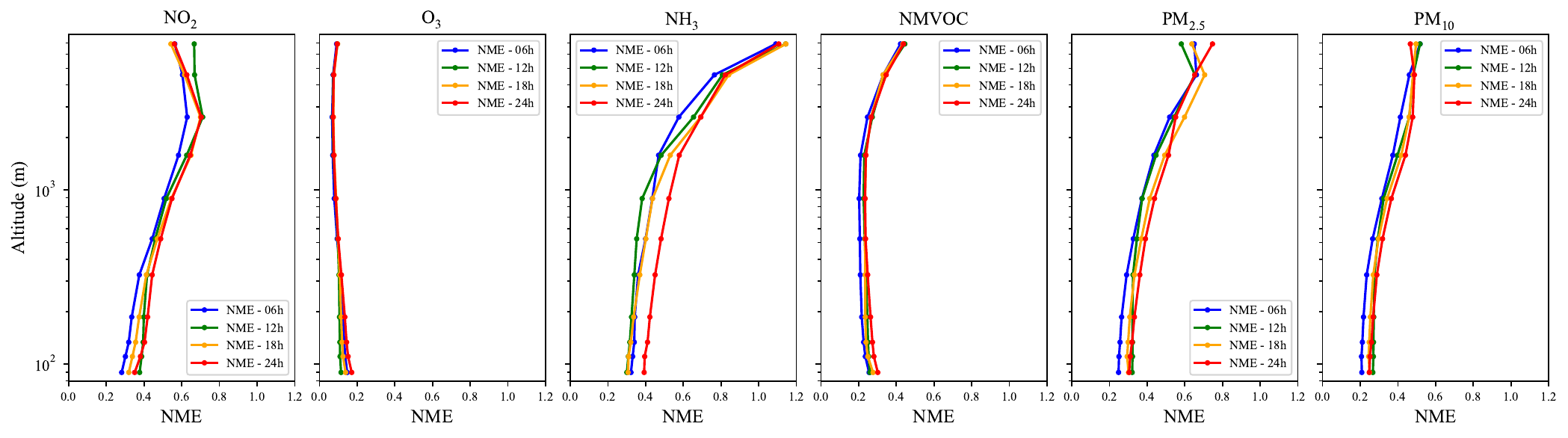}
    \caption{Vertical distribution of NME between \modelname forecast and LOTOS-EUROS for all the variables with the lead time of 6, 12, 18, 24 hours. All the predictions start from 00:00 at each day on 2022.}
    \label{fig:vertical_nme}
\end{figure}

\section{Outliers from simulations}
\label{sect:outlier}
\begin{figure}[ht]
    \centering
    \includegraphics[width=0.9\linewidth]{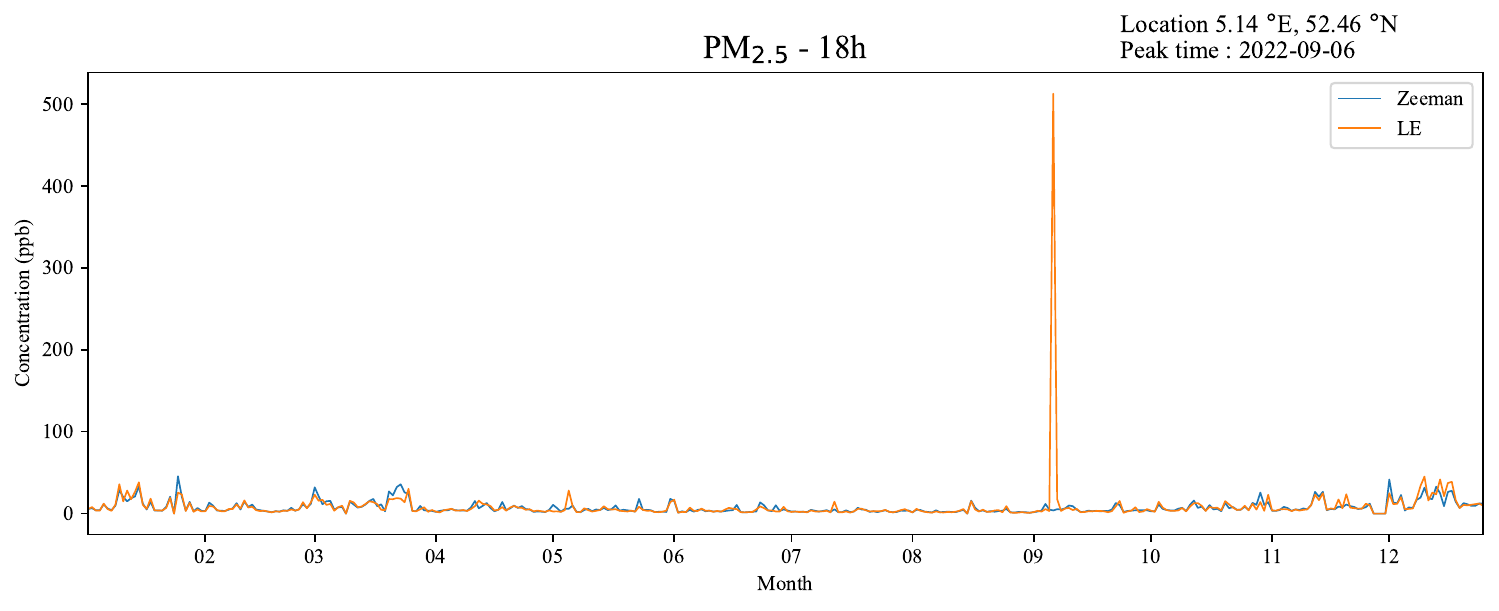}
    \caption{Time series of \modelname forecasts with the lead time of 18h and LOTOS-EUROS simulations at 18:00 on a fix location across the year 2022.}
    \label{fig:conc_series}
\end{figure}

In the LOTOS-EUROS simulations, there are some certain extreme values that caused by the incorrect emissions. It is featured as peak of concentration in a short time. \Cref{fig:conc_series} shows one extreme concentration value at 2022-09-06, location \qty{5.14}{\degree E}, \qty{52.46}{\degree N}. It comes from extreme high fire emissions at this location for the particular days. \modelname is not capturing this pattern since these kinds of emission is irregular in time and location and these emissions are not considered in the training data. 

\section{Forecast series}
\begin{figure}[ht]
    \centering
    \includegraphics[width=0.6\linewidth]{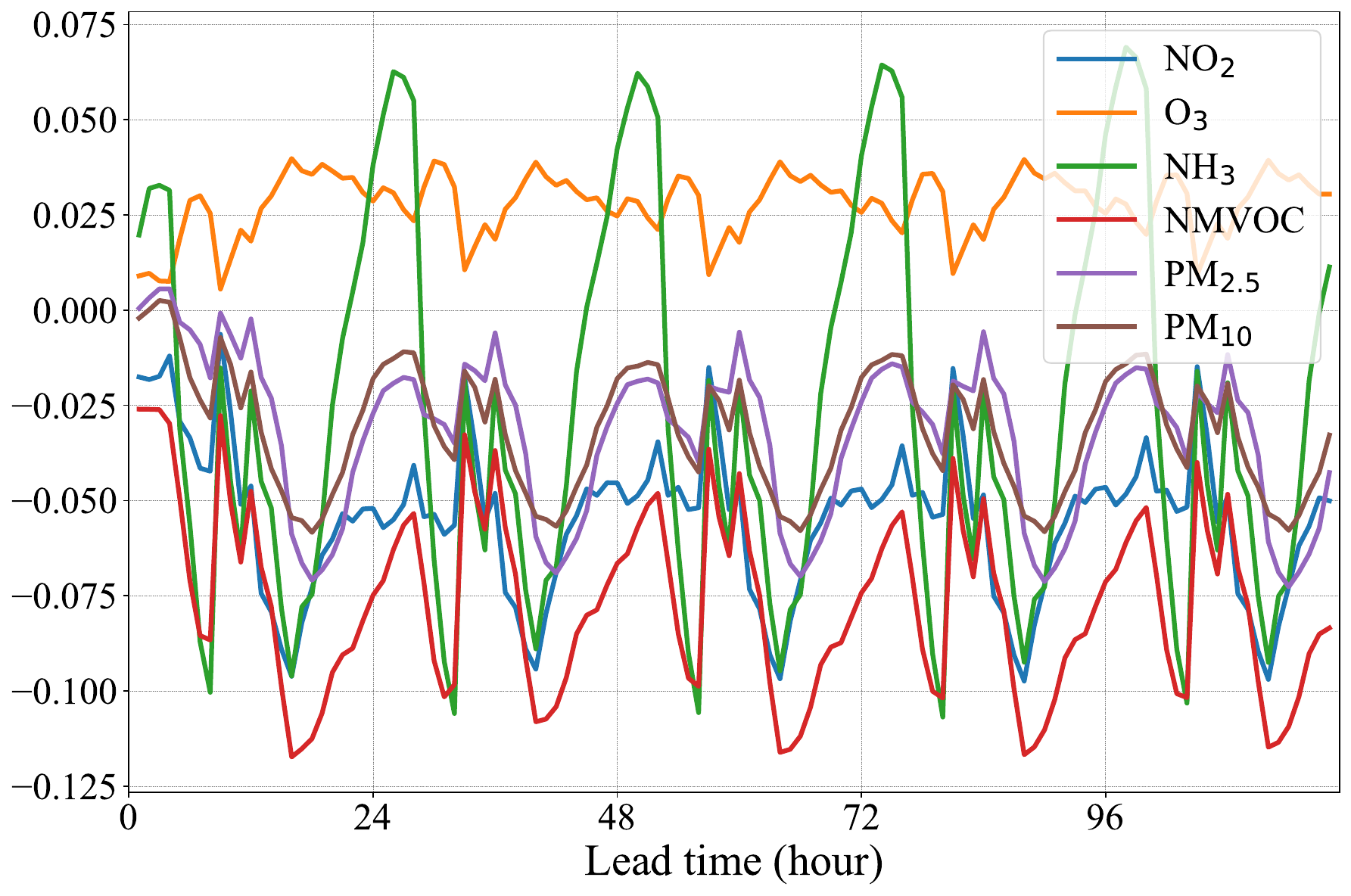}
    \caption{Trend of NMB for the hourly forecasts made by \modelname. The lead time starts from 1 hour to 120 hour. The shown data is on ground level and all the forecasts start from 00:00 at each day on 2022.}
    \label{fig:sequence_nmb}
\end{figure}
\begin{figure}[ht]
    \centering
    \includegraphics[width=0.6\linewidth]{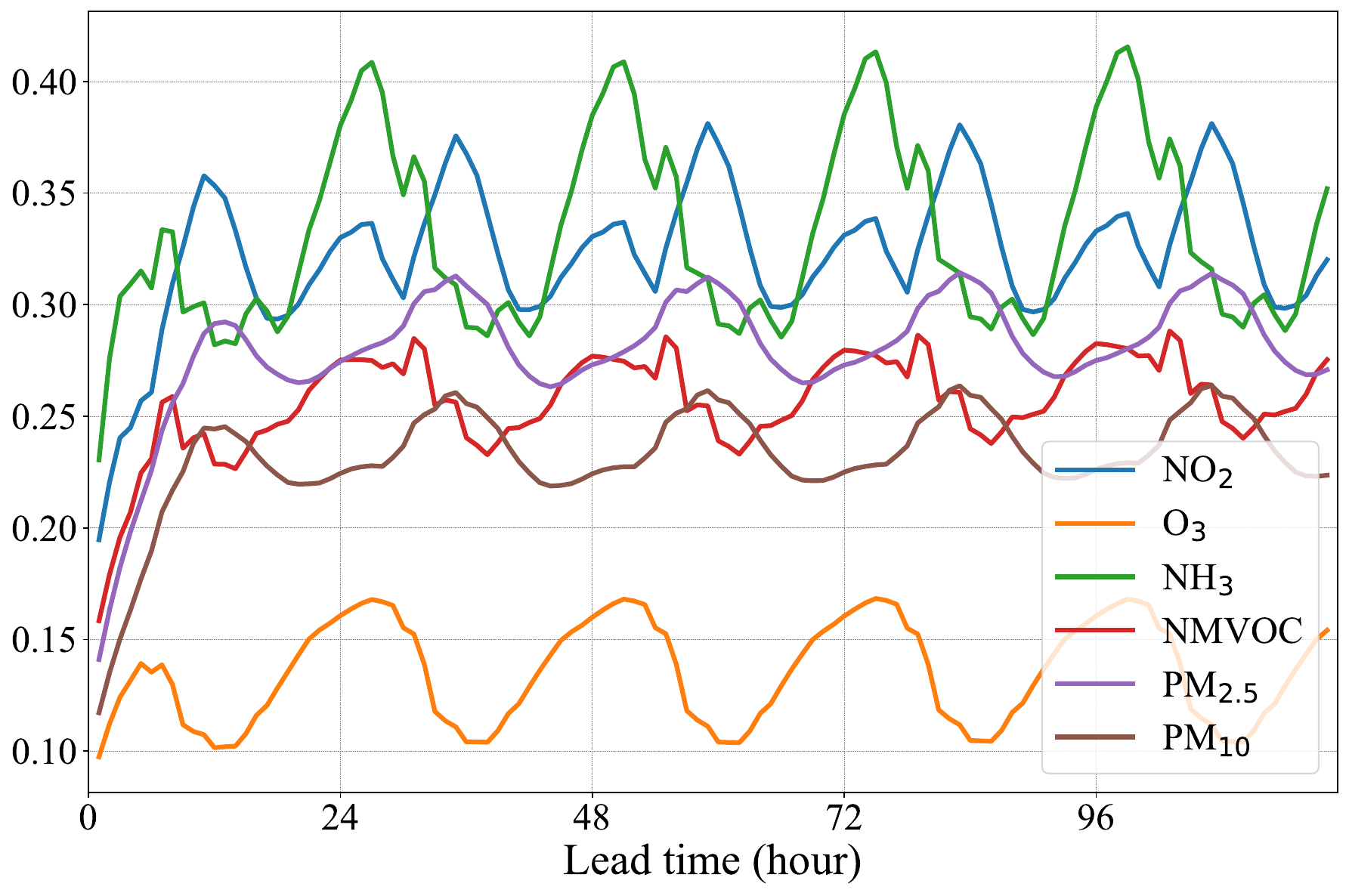}
    \caption{Trend of NME for the hourly forecasts made by \modelname. The lead time starts from 1 hour to 120 hour. The shown data is on ground level and all the forecasts start from 00:00 at each day on 2022.}
    \label{fig:sequence_nme}
\end{figure}
\begin{figure}[ht]
    \centering
    \includegraphics[width=0.6\linewidth]{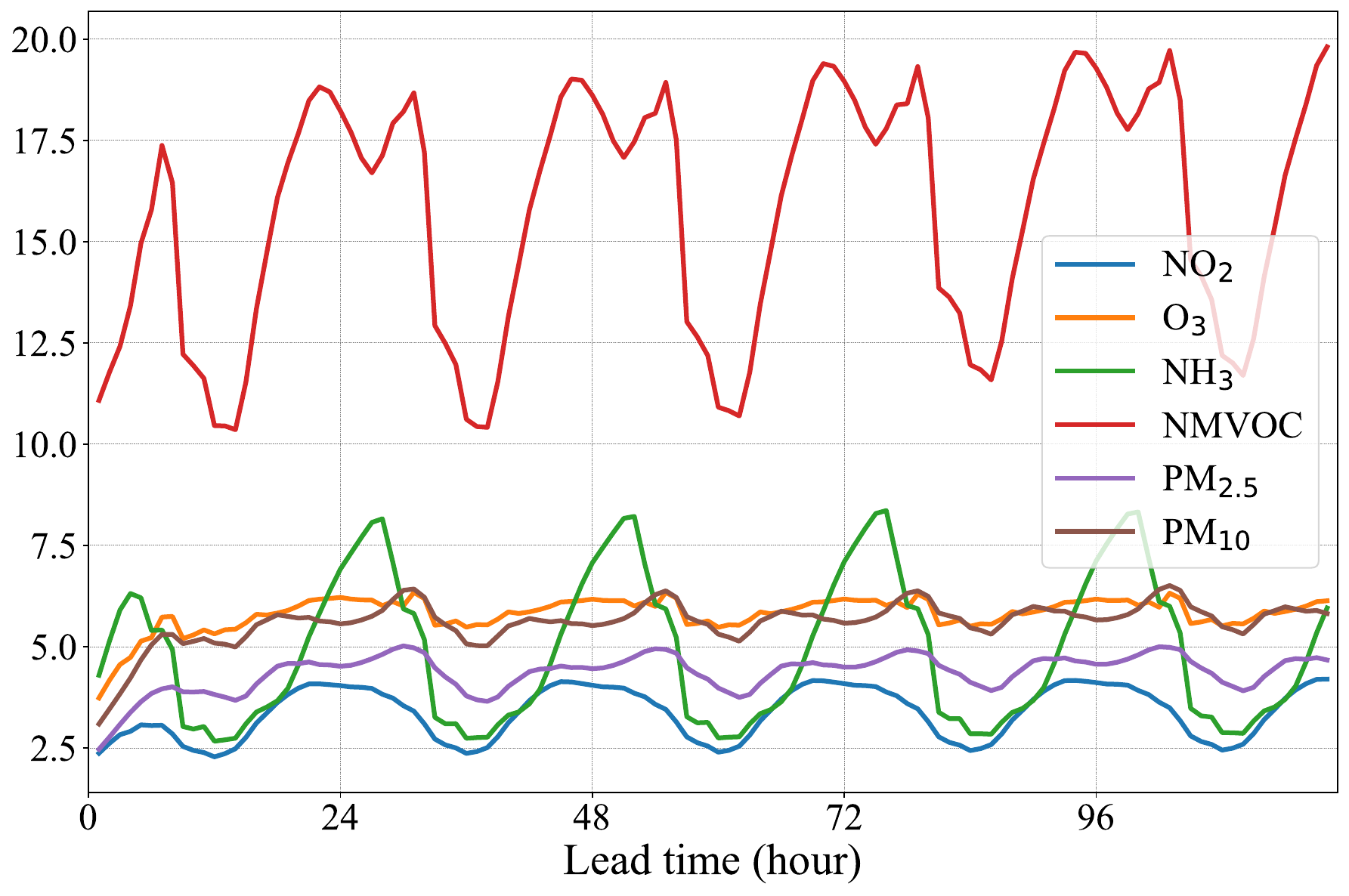}
    \caption{Trend of RMSE for the hourly forecasts made by \modelname. The lead time starts from 1 hour to 120 hour. The shown data is on ground level and all the forecasts start from 00:00 at each day on 2022.}
    \label{fig:sequence_rmse}
\end{figure}

\section{Forecasted series on cities}
\begin{figure}[ht]
    \centering
    \includegraphics[width=0.5\textwidth]{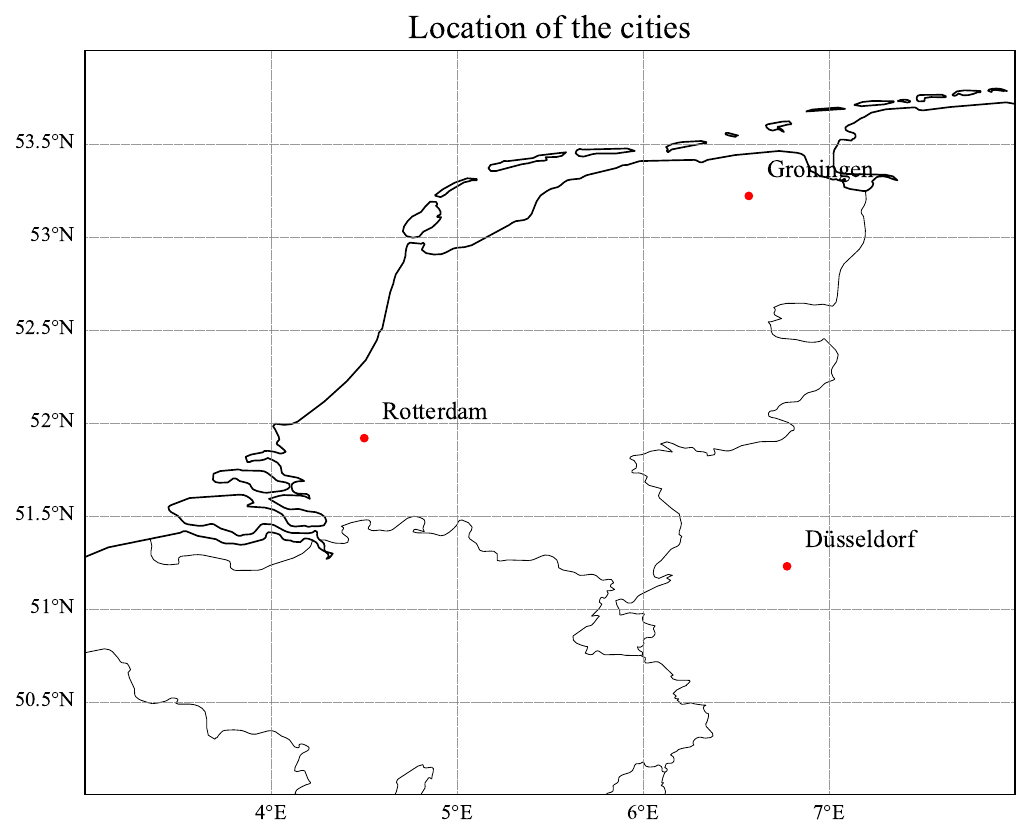}
    \caption{Location of the cities chosen to show the variation of concentrations.}
    \label{fig:city_map}
\end{figure}

Three cities are selected to show the performance of \modelname forecasts in fine scale. Locations of these cities can be seen in \cref{fig:city_map}. Below are the specific coordinates:
\begin{itemize}
    \item Groningen : \qty{6.5675}{\degree E}, \qty{53.2189}{\degree N}
    \item Rotterdam : \qty{4.5}{\degree E}, \qty{53.2189}{\degree N}
    \item Düsseldorf : \qty{6.773056}{\degree E}, \qty{51.227778}{\degree N}
\end{itemize}

\begin{figure}[ht]
    \centering
    \includegraphics[width=1\textwidth]{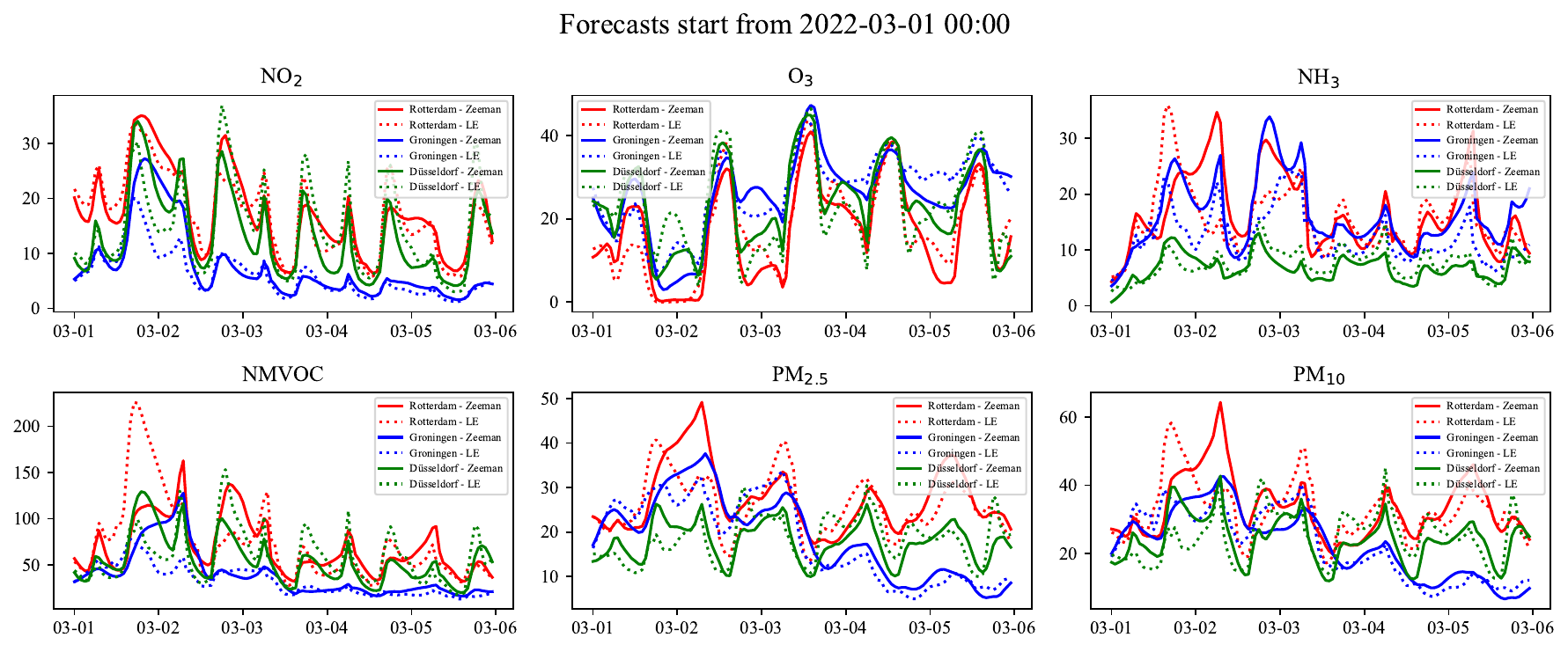}
    \caption{Time series of 5 day forecast on Rotterdam (red), Groningen (blue) and Düsseldorf (green). Dash line is LOTOS-EUROS simulations and solid line is \modelname forecast. The forecast starts from 00:00, 1st March, 2022.}
    \label{fig:point03}
\end{figure}

\begin{figure}[ht]
    \centering
    \includegraphics[width=1\textwidth]{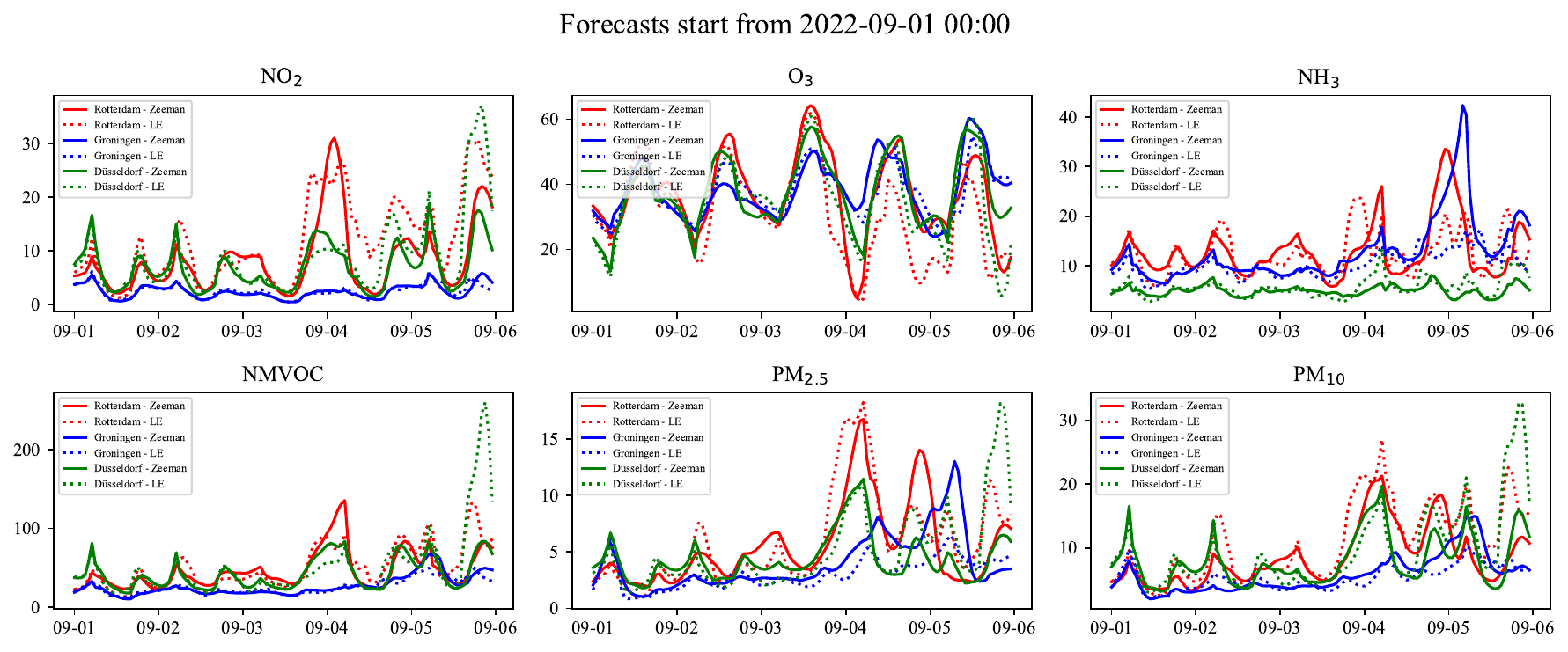}
    \caption{Time series of 5 day forecast on Rotterdam (red), Groningen (blue) and Düsseldorf (green). Dash line is LOTOS-EUROS simulations and solid line is \modelname forecast. The forecast starts from 00:00, 1st September, 2022.}
    \label{fig:point09}
\end{figure}

\begin{figure}[ht]
    \centering
    \includegraphics[width=1\textwidth]{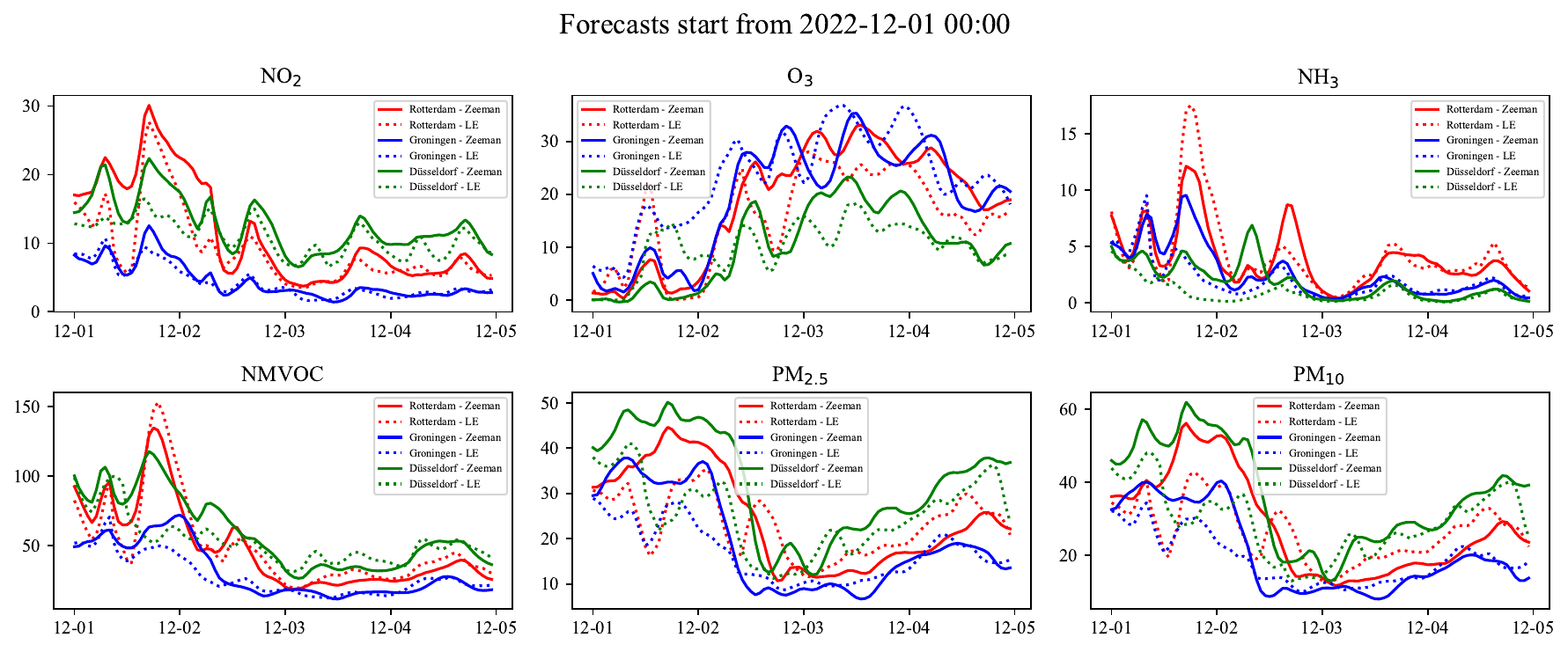}
    \caption{Time series of 5 day forecast on Rotterdam (red), Groningen (blue) and Düsseldorf (green). Dash line is LOTOS-EUROS simulations and solid line is \modelname forecast. The forecast starts from 00:00, 1st December, 2022.}
    \label{fig:point12}
\end{figure}

\end{document}